\documentclass[a4paper,12pt]{article}
\usepackage{epsfig}
\usepackage{graphicx}
\usepackage{amsfonts}
\usepackage{amsmath}
\usepackage{amssymb}
\usepackage{hyperref}
\usepackage{float}
\input{epsf}

\begin{document}

\title{\textbf{Two Curves, One Price:}\\
\emph{Pricing \& Hedging Interest Rate Derivatives }\\
\emph{Decoupling Forwarding and Discounting \\ Yield Curves}}
\author{Marco Bianchetti
\thanks{The author acknowledges fruitful discussions with M. Blatter, M. De Prato, M. Henrard, M. Joshi, C. Maffi, G. V. Mauri, F. Mercurio, N. Moreni, A. Pallavicini, many colleagues in the Risk Management and participants at Quant Congress Europe 2009. A particular mention goes to M. Morini and M. Pucci for their encouragement, and to F. M. Ametrano and the QuantLib community for the open-source developments used here. The views expressed here are those of the author and do not represent the opinions of his employer. They are not responsible for any use that may be made of these contents.}
}
\date{\small{Risk Management, Market Risk, Pricing and Financial Modeling, \\
Intesa Sanpaolo, piazza P. Ferrari 10, 20121 Milan, Italy, \\
e-mail: marco.bianchetti(at)intesasanpaolo.com\\
\vspace{0.1cm}
First version:\ 14 Nov. 2008, this version: 1 Aug. 2012.\\
A shorter version of this paper is published
in Risk Magazine, Aug. 2010.}}
\maketitle

\begin{abstract}
We revisit the problem of pricing and hedging plain vanilla
single-currency interest rate derivatives using multiple distinct yield
curves for market coherent estimation of discount factors and forward rates
with different underlying rate tenors.

Within such double-curve-single-currency framework, adopted by the market
after the credit-crunch crisis started in summer 2007, standard single-curve
no-arbitrage relations are no longer valid, and can be recovered by taking
properly into account the forward basis bootstrapped from market basis
swaps. Numerical results show that the resulting forward basis curves may
display a richer micro-term structure that may induce appreciable effects on
the price of interest rate instruments.

By recurring to the foreign-currency analogy we also derive generalised
no-arbitrage double-curve market-like formulas for basic plain vanilla
interest rate derivatives, FRAs, swaps, caps/floors and swaptions in
particular. These expressions include a quanto adjustment typical of
cross-currency derivatives, naturally originated by the change between the
numeraires associated to the two yield curves, that carries on a volatility
and correlation dependence. Numerical scenarios confirm that such correction
can be non negligible, thus making unadjusted double-curve prices, in
principle, not arbitrage free.

Both the forward basis and the quanto adjustment find a natural financial
explanation in terms of counterparty risk.
\vspace{0.5cm}
\vspace{0.5cm}
\end{abstract}
\vspace{2cm}
\tableofcontents
\newpage

\section{\label{SecIntro}Introduction}

\begin{figure}[hb]
\centering
\includegraphics[scale=0.8]{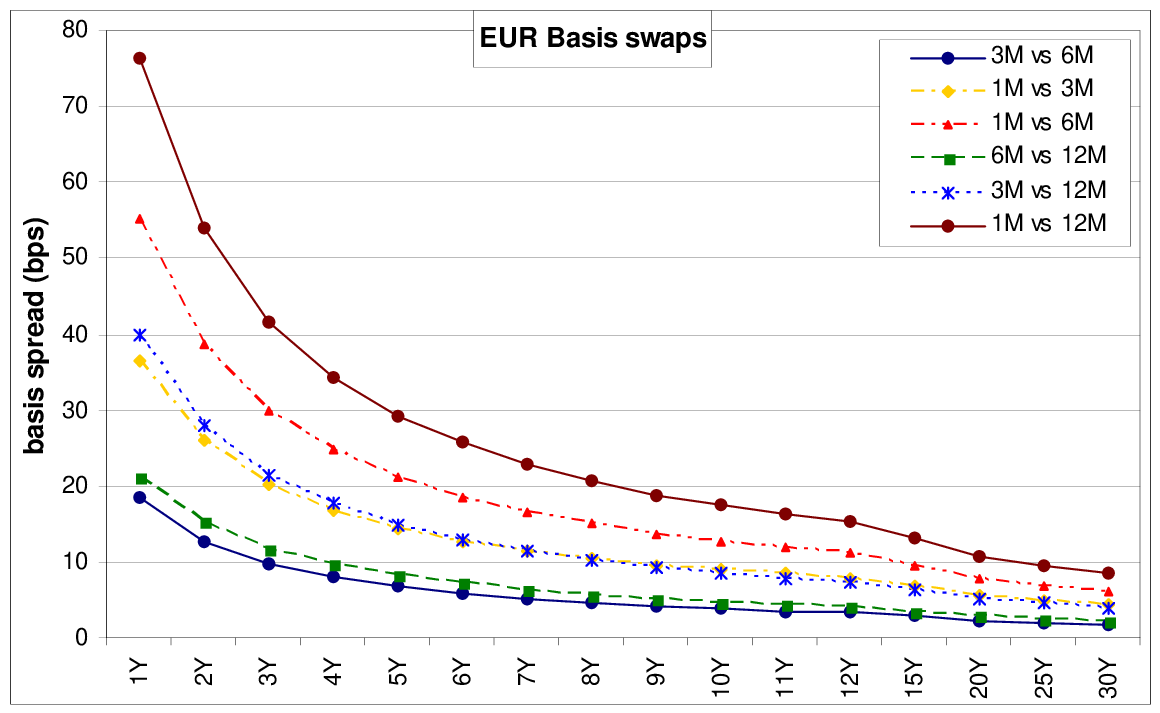}
%\vspace{-0.4cm}
\caption{quotations (basis points) as of Feb. 16th, 2009 for the six EUR basis swap curves corresponding to the four Euribor swap curves 1M, 3M, 6M, 12M. Before the credit crunch of Aug. 2007 the basis spreads were just a few basis points (source:\ Reuters page ICAPEUROBASIS).}
\label{FigBasisSwapReuters}
\end{figure}

The credit crunch crisis started in the second half of 2007 has triggered,
among many consequences, the explosion of the basis spreads quoted on the
market between single-currency interest rate instruments, swaps in
particular, characterised by different underlying rate tenors (e.g. Xibor3M
\footnote{We denote with Xibor a generic Interbank Offered Rate. In the EUR case the
Euribor is defined as the rate at which euro interbank term deposits within
the euro zone are offered by one prime bank to another prime bank (see
\emph{www.euribor.org}).}, Xibor6M, etc.).
In fig. \ref{FigBasisSwapReuters} we show a snapshot of the market quotations as of Feb.
16th, 2009 for the six basis swap term structures corresponding to the four
Euribor tenors 1M, 3M, 6M, 12M. As one can see, in the time interval $1Y-30Y$
the basis spreads are monotonically decreasing from 80 to around 2 basis
points. Such very high basis reflect the higher liquidity risk suffered by
financial institutions and the corresponding preference for receiving
payments with higher frequency (quarterly instead of semi-annually, etc.).
\par
There are also other indicators of regime changes in the interest rate
markets, such as the divergence between deposit (Xibor based) and OIS
\footnote{Overnight Indexed Swaps.} (Eonia\footnote{Euro OverNight Index Average, the rate computed as a weighted average of all overnight rates corresponding to unsecured lending transactions in the euro-zone interbank market (see e.g. \emph{www.euribor.org}).} based for
EUR) rates, or between FRA\footnote{Forward Rate Agreement.} contracts and the corresponding forward rates implied by consecutive deposits (see e.g. refs. \cite{AmeBia09}, \cite{Mer09}, \cite{Mor08}, \cite{Mor09}).

These frictions have thus induced a sort of \textquotedblleft
segmentation\textquotedblright\ of the interest rate market into sub-areas,
mainly corresponding to instruments with 1M, 3M, 6M, 12M underlying rate
tenors, characterized, in principle, by different internal dynamics,
liquidity and credit risk premia, reflecting the different views and
interests of the market players. We stress that market segmentation was already
present (and well understood) before the credit crunch (see e.g. ref. \cite%
{TucPor03}), but not effective due to negligible basis spreads.

Such evolution of the financial markets has triggered a general reflection
about the methodology used to price and hedge interest rate derivatives,
namely those financial instruments whose price depends on the present value
of future interest rate-linked cash flows. In this paper we acknowledge the
current market practice, assuming the existence of a given methodology
(discussed in detail in ref. \cite{AmeBia09}) for bootstrapping multiple
homogeneous forwarding and discounting curves, characterized by different
underlying rate tenors, and we focus on the consequences for pricing and
hedging interest rate derivatives. In particular in sec. \ref{SecFrameworks}
we summarise the pre- and post-credit crunch market practices for pricing and
hedging interest rate derivatives. In sec. \ref{SecNotation} we fix
the notation, we revisit some general concept of standard, no arbitrage
single-curve pricing and we formalize the double-curve pricing framework,
showing how no arbitrage is broken and can be formally recovered with the
introduction of a forward basis. In sec. \ref{SecQuantoAdj} we use the
foreign-currency analogy to derive a single-currency version of the quanto
adjustment, typical of cross-currency derivatives, naturally appearing in
the expectation of forward rates. In sec. \ref{SecPricing} we derive the no
arbitrage double-curve market-like pricing expressions for basic
single-currency interest rate derivatives, such as FRA, swaps, caps/floors
and swaptions. Conclusions are summarised in sec. \ref{SecConclusions}.
\par
The topic discussed here is a central problem in the interest rate market,
with many consequences in trading, financial control, risk management and
IT, which still lacks of attention in the financial literature. To our
knowledge, similar topics have been approached in refs. \cite{FruZam1995},
\cite{BoeSch2005}, \cite{KijTan08}, \cite{Mer09}, \cite{Hen09}and \cite%
{Mor08}, \cite{Mor09} . In particular W. Boenkost and W. Schmidt \cite%
{BoeSch2005} discuss two methodologies for pricing cross-currency basis
swaps, the first of which (the actual pre-crisis common market practice),
does coincide, once reduced to the single-currency case, with the
double-curve pricing procedure described here\footnote{%
these authors were puzzled by the fact that their first methodology was
neither arbitrage free nor consistent with the pre-crisis single-curve
market practice for pricing single-currency swaps. Such objections have now
been overcome by the market evolution towards a generalized double-curve
pricing approach (see also \cite{TucPor03}).}. Recently M. Kijima et al.
\cite{KijTan08} have extended the approach of ref. \cite{BoeSch2005} to the
(cross currency) case of three curves for discount rates, Libor rates and
bond rates. Finally, simultaneously to the development of the present paper,
M. Morini is approaching the problem in terms of counterparty risk \cite%
{Mor08}, \cite{Mor09}, F. Mercurio in terms of an extended Libor Market
Model \cite{Mer09}, and M. Henrard using an axiomatic model \cite{Hen09}.

The present work follows an alternative route with respect to those cited
above, in the sense that a) we adopt a \emph{bottom-up} practitioner's
perspective, starting from the current market practice of using multiple
yield curves and working out its natural consequences, looking for a minimal
and light generalisation of well-known frameworks, keeping things as simple
as possible; b) we show how no-arbitrage can be recovered in the
double-curve approach by taking properly into account the forward basis,
whose term structure can be extracted from available basis swap market
quotations; c) we use a straightforward foreign-currency analogy to derive
generalised double-curve market-like pricing expressions for basic
single-currency interest rate derivatives, such as FRAs, swaps, caps/floors
and swaptions.

\section{\label{SecNotation}Notation and Basic Assumptions}

Following the discussion above, we denote with $M_{x}$, $x=\left\{
d,f_{1},...,f_{n}\right\} $ multiple distinct interest rate sub-markets,
characterized by the same currency and by distinct bank accounts $B_{x}$,
such that
\begin{equation}
B_x(t)=\exp{\int_0^t r_x(u)\,du},
\end{equation}
where $r_x(t)$ are the associated short rates.
We also have multiple yield curves $\complement_{x}$ in the form of a continuous term structure
of discount factors,
\begin{equation}
\complement_{x}=\left\{ T\longrightarrow P_{x}\left( t_{0},T\right) ,T\geq t_{0}\right\},
\end{equation}
where $t_{0}$\ is the reference date of the curves (e.g. settlement date, or
today) and $P_{x}\left( t,T\right) $ denotes the price at time $t\geq t_{0}$
of the $M_{x}$-zero coupon bond for maturity $T$, such that $P_{x}\left(
T,T\right) =1$. In each sub-market $M_{x}$ we postulate the usual no
arbitrage relation,
\begin{equation}
P_{x}\left( t,T_{2}\right) =P_{x}\left( t,T_{1}\right) P_{x}\left(t,T_{1},T_{2}\right),
\;t\leq T_{1}<T_{2},
\label{NoArbitrage}
\end{equation}
where $P_{x}\left( t,T_{1},T_{2}\right) $ denotes the $M_{x}$ forward
discount factor from time $T_{2}$ to time $T_{1}$, prevailing at time $t$.
The financial meaning of expression \ref{NoArbitrage} is that, in each
market $M_{x}$, given a cash flow of one unit of currency at time $T_{2}$,
its corresponding value at time $t<T_{2}$ must be unique, both if we
discount in one single step from $T_{2}$ to $t$, using the discount factor
$P_{x}\left( t,T_{2}\right) $, and if we discount in two steps, first from
$T_{2}$ to $T_{1}$, using the forward discount $P_{x}\left(
t,T_{1},T_{2}\right) $ and then from $T_{1}$ to $t$, using $P_{x}\left(
t,T_{1}\right) $. Denoting with $F_{x}\left( t;T_{1},T_{2}\right) $ the
simple compounded forward rate associated to $P_{x}\left(
t,T_{1},T_{2}\right) $, resetting at time $T_{1}$ and covering the time
interval $\left[ T_{1};T_{2}\right] $, we have%
\begin{equation}
P_{x}\left( t,T_{1},T_{2}\right)
=\frac{P_{x}\left( t,T_{2}\right) }{P_{x}\left( t,T_{1}\right)} =\frac{1}{1+F_{x}\left(t;T_{1},T_{2}\right)
\tau_{x}\left( T_{1},T_{2}\right) },
\end{equation}%
where $\tau_{x}\left( T_{1},T_{2}\right) $ is the year fraction between
times $T_{1}$ and $T_{2}$ with daycount $dc_{x}$, and from eq. \ref{NoArbitrage} we obtain the familiar no arbitrage expression
\begin{eqnarray}
F_{x}\left( t;T_{1},T_{2}\right) &=&\frac{1}{\tau_{x}\left(
T_{1},T_{2}\right) }\left[ \frac{1}{P_{x}\left( t,T_{1},T_{2}\right) }-1\right]  \notag \\
&=&\frac{P_{x}\left( t,T_{1}\right) -P_{x}\left( t,T_{2}\right) }
{\tau_{x}\left( T_{1},T_{2}\right) P_{x}\left( t,T_{2}\right) }.
\label{FwdRate}
\end{eqnarray}
Eq. \ref{FwdRate} can be also derived (see e.g. ref. \cite{BriMer06}, sec.
1.4) as the fair value condition at time $t$ of the Forward Rate Agreement
(FRA) contract with payoff at maturity $T_{2}$ given by%
\begin{gather}
\text{\textbf{FRA}}_{x}\left( T_{2};T_{1},T_{2},K,N\right)
=N\tau_{x}\left(T_{1},T_{2}\right) \left[ L_{x}\left( T_{1},T_{2}\right) -K\right] ,
\label{FRAPayoff} \\
L_{x}\left( T_{1},T_{2}\right)
=\frac{1-P_{x}\left( T_{1},T_{2}\right) }{\tau_{x}\left( T_{1},T_{2}\right)
P_{x}\left( T_{1},T_{2}\right) }
\end{gather}
where $N$ is the nominal amount, $L_{x}\left( T_{1},T_{2},dc_{x}\right) $ is
the $T_{1}$-spot Xibor rate for maturity $T_{2}$ and $K$ the (simply
compounded) strike rate (sharing the same daycount convention for
simplicity). Introducing expectations we have, $\forall t\leq T_{1}<T_{2}$,
\begin{align}
\text{\textbf{FRA}}_{x}\left( t;T_{1},T_{2},K,N\right) & =P_{x}\left(
t,T_{2}\right) \mathbb{E}_{t}^{Q_{x}^{T_{2}}}\left[ \text{\textbf{FRA}}
\left( T_{2};T_{1},T_{2},K,N\right) \right]  \notag \\
& =NP_{x}\left( t,T_{2}\right) \tau_{x}\left( T_{1},T_{2}\right) \left\{
\mathbb{E}_{t}^{Q_{x}^{T_{2}}}\left[ L_{x}\left( T_{1},T_{2}\right) \right]
-K\right\}  \notag \\
& =NP_{x}\left( t,T_{2}\right) \tau_{x}\left( T_{1},T_{2}\right) \left[
F_{x}\left( t;T_{1},T_{2}\right) -K\right] ,  \label{FRAPrice}
\end{align}%
where $Q_{x}^{T_{2}}$ denotes the $M_{x}$-$T_{2}$-forward measure
corresponding to the numeraire
$P_{x}\left( t,T_{2}\right) $, $\mathbb{E}_{t}^{Q}\left[ .\right] $
denotes the expectation at time $t$ w.r.t. measure
$Q$ and filtration $\mathcal{F}_{t}$, encoding the market information
available up to time $t$, and we have assumed the standard martingale
property of forward rates%
\begin{equation}
F_{x}\left( t;T_{1},T_{2}\right) =\mathbb{E}_{t}^{Q_{x}^{T_{2}}}\left[
F_{x}\left( T_{1};T_{1},T_{2}\right) \right] =\mathbb{E}_{t}^{Q_{x}^{T_{2}}}
\left[ L_{x}\left( T_{1},T_{2}\right) \right]
\end{equation}%
to hold in each interest rate market $M_{x}$ (see e.g. ref. \cite{BriMer06}).
We stress that the assumptions above imply that \emph{each} sub-market $M_{x}$
is internally consistent and has the same properties of the "classical" interest rate market before the
crisis. This is surely a strong hypothesis, that could be relaxed in more sophisticated frameworks.

\section{\label{SecFrameworks}Pre and Post Credit Crunch Market Practices}
We describe here the evolution of the market practice for pricing and hedging interest rate derivatives through the credit crunch crisis.
We use consistently the notation described above, considering a general single-currency interest rate derivative with $m$ future coupons with payoffs
$\mathbf{\pi }=\left\{ \pi _{1},...,\pi _{m}\right\}$,
with $\pi _{i}=\pi _{i}\left( F_{x}\right) $, generating $m$ cash flows
$\mathbf{c}=\left\{ c_{1},...,c_{m}\right\} $ at future dates
$\mathbf{T}=\left\{ T_{1},...,T_{m}\right\} $, with $t<T_{1}<...<T_{m}$.

\subsection{\label{SecSingleCurve}Single-Curve Framework}

The pre-crisis standard market practice was based on a single-curve procedure, well known to the financial world, that can be summarised as follows (see e.g. refs. \cite{Ron00}, \cite{HagWes06}, \cite{And07} and \cite{HagWes08}):

\begin{enumerate}
\item select a \emph{single} finite set of the most convenient (i.e. liquid) interest rate vanilla instruments traded on the market with increasing maturities and build a \emph{single} yield curve $\complement_{d}$ using the preferred bootstrapping procedure (pillars, priorities, interpolation, etc.); for instance, a common choice in the EUR\ market is a combination of short term EUR\ deposits, medium-term Futures/FRA on Euribor3M and medium/long term swaps on Euribor6M;

\item for each interest rate coupon $i\in \left\{ 1,...,m\right\}$ compute the relevant forward rates using the given yield curve $\complement_{d}$ as in eq. \ref{FwdRate},
    \begin{equation}
    F_{d}\left( t;T_{i-1},T_{i}\right) =\frac{P_{d}\left( t,T_{i-1}\right)
    -P_{d}\left( t,T_{i}\right) }{\tau_{d}\left( T_{i-1},T_{i}\right)
    P_{d}\left( t,T_{i}\right) }\;t\leq T_{i-1}<T_{i};
    \end{equation}

\item compute cash flows $c_{i}$ as expectations at time $t$ of the corresponding coupon payoffs $\pi _{i}\left( F_{d}\right) $ with respect to the $T_{i}$-forward measure $Q_{d}^{T_{i}}$, associated to the numeraire $P_{d}\left( t,T_{i}\right)$ from the \emph{same} yield curve $\complement_{d}$,
    \begin{equation}
    c_{i}=c\left( t,T_{i},\pi _{i}\right) =\mathbb{E}_{t}^{Q_{d}^{T_{i}}}\left[
    \pi _{i}\left( F_{d}\right) \right] ;
    \end{equation}

\item compute the relevant discount factors $P_{d}\left( t,T_{i}\right) $ from the \emph{same} yield curve $\complement_{d}$;

\item compute the derivative's price at time $t$ as the sum of the discounted cash flows,
    \begin{equation}
    \pi \left( t;\mathbf{T}\right)
    =\sum_{i=1}^{m}P_{d}\left( t,T_{i}\right)c\left( t,T_{i},\pi _{i}\right)
    =\sum_{i=1}^{m}P_{d}\left( t,T_{i}\right)
    \mathbb{E}_{t}^{Q_{d}^{T_{i}}}\left[ \pi _{i}\left( F_{d}\right) \right] ;
    \end{equation}

\item compute the delta sensitivity with respect to the market pillars of yield curve $\complement_{d}$ and hedge the resulting delta risk using the suggested amounts (hedge ratios) of the \emph{same} set of vanillas.
\end{enumerate}

For instance, a 5.5Y maturity EUR floating swap leg on Euribor1M (not
directly quoted on the market)\ is commonly priced using discount factors
and forward rates calculated on the same depo-Futures-swap curve cited
above. The corresponding delta risk is hedged using the suggested amounts
(hedge ratios)\ of 5Y and 6Y Euribor6M swaps\footnote{%
we refer here to the case of local yield curve bootstrapping methods, for
which there is no sensitivity delocalization effect (see refs. \cite%
{HagWes06}, \cite{HagWes08}).}.

Notice that step 3 above has been formulated in terms of the pricing measure
$Q_{d}^{T_{i}}$ associated to the numeraire $P_{d}\left( t,T_{i}\right) $.
This is convenient in our context because it emphasizes that the numeraire
is associated to the discounting curve. Obviously any other equivalent
measure associated to different numeraires may be used as well.

We stress that this is a \emph{single-currency-single-curve approach}, in
that a \emph{unique} yield curve is built and used to price and hedge any
interest rate derivative on a given currency. Thinking in terms of more
fundamental variables, e.g. the short rate, this is equivalent to assume
that there exist a unique fundamental underlying short rate process able to
model and explain the whole term structure of interest rates of all tenors.
It is also a \emph{relative pricing} approach, because both the price and
the hedge of a derivative are calculated relatively to a set of vanillas
quoted on the market. We notice also that it is not strictly guaranteed to
be arbitrage-free, because discount factors and forward rates obtained from
a given yield curve through interpolation are, in general, not necessarily
consistent with those obtained by a no arbitrage model; in practice bid-ask
spreads and transaction costs hide any arbitrage possibilities. Finally, we
stress that the key first point in the procedure is much more a matter of
art than of science, because there is not an unique financially sound recipe
for selecting the bootstrapping instruments and rules.

\subsection{\label{SecMultipleCurve}Multiple-Curve Framework}

Unfortunately, the pre-crisis approach outlined above is no longer
consistent, at least in its simple formulation, with the present market
conditions. First, it does not take into account the market information
carried by basis swap spreads, now much larger than in the past and no
longer negligible. Second, it does not take into account that the interest
rate market is segmented into sub-areas corresponding to instruments with
distinct underlying rate tenors, characterized, in principle, by \emph{%
different} dynamics (e.g. short rate processes). Thus, pricing and hedging
an interest rate derivative on a single yield curve mixing different
underlying rate tenors can lead to \textquotedblleft
dirty\textquotedblright\ results, incorporating the different dynamics, and
eventually the inconsistencies, of distinct market areas, making prices and
hedge ratios less stable and more difficult to interpret. On the other side,
the more the vanillas and the derivative share the same homogeneous
underlying rate, the better should be the relative pricing and the hedging.
Third, by no arbitrage, discounting must be unique: two identical future
cash flows of whatever origin must display the \emph{same} present value;
hence we need a unique discounting curve.

In principle, a consistent credit and liquidity theory would be required to
account for the interest rate market segmentation. This would also explain
the reason why the asymmetries cited above do not necessarily lead to
arbitrage opportunities, once counterparty and liquidity risks are taken
into account. Unfortunately such a framework is not easy to construct (see
e.g. the discussion in refs. \cite{Mer09}, \cite{Mor09}). In practice an
empirical approach has prevailed on the market, based on the construction of
multiple \textquotedblleft forwarding\textquotedblright\ yield curves from
plain vanilla market instruments \emph{homogeneous} in the underlying rate
tenor, used to calculate future cash flows based on forward interest rates
with the corresponding tenor, and of a single \textquotedblleft
discounting\textquotedblright\ yield curve, used to calculate discount
factors and cash flows' present values. Consequently, interest rate
derivatives with a given underlying rate tenor should be priced and hedged
using vanilla interest rate market instruments with the \emph{same}
underlying rate tenor. The post-crisis market practice may thus be
summarised in the following working procedure:

\begin{enumerate}
\item build \emph{one discounting curve} $\complement_{d}$ using the
preferred selection of vanilla interest rate market instruments and
bootstrapping procedure;

\item build \emph{multiple distinct forwarding curves} $\complement_{f_{1}},...,\complement_{f_{n}}$ using the preferred selections of
distinct sets of vanilla interest rate market instruments, each \emph{%
homogeneous} in the underlying Xibor rate tenor (typically with 1M, 3M, 6M,
12M tenors) and bootstrapping procedures;

\item for each interest rate coupon $i\in \left\{ 1,...,m\right\} $ compute
the relevant forward rates with tenor $f$ using the \emph{corresponding}
yield curve $\complement_{f}$ as in eq. \ref{FwdRate},%
\begin{equation}
F_{f}\left( t;T_{i-1},T_{i}\right) =\frac{P_{f}\left( t,T_{i-1}\right)
-P_{f}\left( t,T_{i}\right) }{\tau_{f}\left( T_{i-1},T_{i}\right)
P_{f}\left( t,T_{i}\right) },\;\;t\leq T_{i-1}<T_{i};
\end{equation}

\item compute cash flows $c_{i}$ as expectations at time $t$ of the
corresponding coupon payoffs $\pi _{i}\left( F_{f}\right) $ with respect to
the \emph{discounting} $T_{i}$-forward measure $Q_{d}^{T_{i}}$, associated
to the numeraire $P_{d}\left( t,T_{i}\right) $, as%
\begin{equation}
c_{i}=c\left( t,T_{i},\pi _{i}\right) =\mathbb{E}_{t}^{Q_{d}^{T_{i}}}\left[
\pi _{i}\left( F_{f}\right) \right] ;  \label{Cashflow}
\end{equation}

\item compute the relevant discount factors $P_{d}\left( t,T_{i}\right) $
from the \emph{discounting} yield curve $\complement_{d}$;

\item compute the derivative's price at time $t$ as the sum of the
discounted cash flows,
\begin{equation}
\pi \left( t;\mathbf{T}\right) =\sum_{i=1}^{m}P_{d}\left( t,T_{i}\right)
c\left( t,T_{i},\pi _{i}\right) =\sum_{i=1}^{m}P_{d}\left( t,T_{i}\right)
\mathbb{E}_{t}^{Q_{d}^{T_{i}}}\left[ \pi _{i}\left( F_{f}\right) \right] ;
\label{Price}
\end{equation}

\item compute the delta sensitivity with respect to the market pillars of \emph{each }yield curve $\complement_{d},\complement_{f_{1}},...,\complement_{f_{n}}$ and hedge the resulting delta risk using the suggested amounts (hedge ratios) of the \emph{corresponding} set of
vanillas.
\end{enumerate}

For instance, the 5.5Y floating swap leg cited in the previous section \ref{SecSingleCurve} is currently priced using Euribor1M forward rates
calculated on the $\complement_{1M}$ forwarding curve, bootstrapped using
Euribor1M vanillas only, plus discount factors calculated on the discounting
curve $\complement_{d}$. The delta sensitivity is computed by shocking one
by one the market pillars of \emph{both} $\complement_{1M}$ and $%
\complement_{d}$ curves and the resulting delta risk is hedged using the
suggested amounts (hedge ratios) of 5Y and 6Y Euribor1M swaps plus the
suggested amounts of 5Y and 6Y instruments from the discounting curve $%
\complement_{d}$ (see sec. \ref{SecHedging} for more details about the
hedging procedure).

Such multiple-curve framework is consistent with the present market
situation, but - there is no free lunch - it is also more demanding. First,
the discounting curve clearly plays a special and fundamental role, and must
be built with particular care. This \textquotedblleft
pre-crisis\textquotedblright\ obvious step has become, in the present market
situation, a very subtle and controversial point, that would require a whole
paper in itself (see e.g. ref. \cite{Hen07}). In fact, while the forwarding
curves construction is driven by the underlying rate homogeneity principle,
for which there is (now) a general market consensus, there is no longer, at
the moment, general consensus for the discounting curve construction. At
least two different practices can be encountered in the market:
a) the old \textquotedblleft pre-crisis\textquotedblright\ approach (e.g. the depo,
Futures/FRA and swap curve cited before), that can be justified with the principle of maximum liquidity (plus a little of inertia), and
b) the OIS curve, based on the overnight rate (Eonia for EUR), considered as the best proxy to a risk free rate available on the market because of its 1-day tenor, justified with collateralized (riskless) counterparties
\footnote{collateral agreements are more and more used in OTC markets, where there are
no clearing houses,\ to reduce the counterparty risk. The standard ISDA
contracts (ISDA\ Master Agreement and Credit Support Annex) include netting
clauses imposing compensation. The compensation frequency is often on a
daily basis and the (cash or asset) compensation amount is remunerated at
overnight rate.} (see e.g. refs. \cite{Mad08}, \cite{GS09}).
Second, building multiple curves requires multiple quotations: many more
bootstrapping instruments must be considered (deposits, Futures, swaps,
basis swaps, FRAs, etc., on different underlying rate tenors), which are
available on the market with different degrees of liquidity and can display
transitory inconsistencies (see \cite{AmeBia09}). Third, non trivial
interpolation algorithms are crucial to produce smooth forward curves (see
e.g. refs. \cite{HagWes06}-\cite{HagWes08}, \cite{AmeBia09}). Fourth,
multiple bootstrapping instruments implies multiple sensitivities, so
hedging becomes more complicated. Last but not least, pricing libraries,
platforms, reports, etc. must be extended, configured, tested and released
to manage multiple and separated yield curves for forwarding and
discounting, not a trivial task for quants, risk managers, developers and
IT\ people.

The static multiple-curve pricing \& hedging methodology described above can
be extended, in principle, by adopting multiple distinct models for the
evolution of the underlying interest rates with tenors $f_{1},...,f_{n}$ to
calculate the future dynamics of the yield curves $\complement_{f_{1}},...,\complement_{f_{n}}$ and the expected cash flows. The volatility/correlation dependencies carried by such models imply, in principle, bootstrapping multiple distinct variance/covariance matrices and
hedging the corresponding sensitivities using volatility- and
correlation-dependent vanilla market instruments. Such more general approach
has been carried on in ref. \cite{Mer09} in the context of generalised
market models. In this paper we will focus only on the basic matter of
static yield curves and leave out the dynamical volatility/correlation
dimensions.

\section{\label{SecNoArbBasisAdj}No Arbitrage and Forward Basis}

Now, we wish to understand the consequences of the assumptions above in
terms of no arbitrage. First, we notice that, in the multiple-curve framework,
classical single-curve no arbitrage relations are broken. 
For instance, if we assign index $d$ to discount factors and index $f$ to forward discount factors (containing forward rates) in eqs. \ref{NoArbitrage} and \ref{FwdRate}, we obtain
\begin{gather}
%P_d\left(t,T_2\right) = P_d\left(t,T_1\right) P_f\left(t,T_1,T_2\right),\\
P_f\left(t,T_1,T_2\right) = \frac{P_d\left(t,T_2\right)}{P_d\left(t,T_1\right)} ,\\
P_f\left(t,T_1,T_2\right)
=\frac{1}{1+F_{f}\left(t;T_{1},T_{2}\right) \tau_{f}\left( T_{1},T_{2}\right) }
=\frac{P_f(t,T_2)}{P_f(t,T_1)},
\end{gather}
which are clearly inconsistent.
No arbitrage between distinct yield curves $\complement_{d}$ and $\complement_{f}$ can be immediately recovered by taking into account the
\emph{forward basis}, the forward counterparty of the quoted market basis
of fig. \ref{FigBasisSwapReuters}, defined as%
\begin{equation}
P_{f}\left( t,T_{1},T_{2}\right) :=\frac{1}{1+F_{d}\left(
t;T_{1},T_{2}\right) BA_{fd}\left( t,T_{1},T_{2}\right) \tau_{d}\left(
T_{1},T_{2}\right) },  \label{NoArbitrage2}
\end{equation}
or through the equivalent simple transformation rule for forward rates
\begin{equation}
F_{f}\left( t;T_{1},T_{2}\right) \tau_{f}\left( T_{1},T_{2}\right)
=F_{d}\left( t;T_{1},T_{2}\right) \tau_{d}\left( T_{1},T_{2}\right)
BA_{fd}\left( t,T_{1},T_{2}\right) .  \label{FwdBasisAdj}
\end{equation}%
From eq. \ref{FwdBasisAdj} we can express the forward basis as a ratio
between forward rates or, equivalently, in terms of discount factors from $\complement_{d}$ and $\complement_{f}$ curves as
\begin{eqnarray}
BA_{fd}\left( t,T_{1},T_{2}\right) &=&\frac{F_{f}\left( t;T_{1},T_{2}\right)
\tau_{f}\left( T_{1},T_{2}\right) }{F_{d}\left( t;T_{1},T_{2}\right) \tau_{d}\left( T_{1},T_{2}\right) }  \notag \\
&=&\frac{P_{d}\left( t,T_{2}\right) }{P_{f}\left( t,T_{2}\right) }\frac{%
P_{f}\left( t,T_{1}\right) -P_{f}\left( t,T_{2}\right) }{P_{d}\left(
t,T_{1}\right) -P_{d}\left( t,T_{2}\right) }.  \label{FwdBasisAdj2}
\end{eqnarray}
Obviously the following alternative additive definition is completely equivalent
\begin{equation}
P_{f}\left( t,T_{1},T_{2}\right)
:=\frac{1}{1+\left[ F_{d}\left(t;T_{1},T_{2}\right) +BA_{fd}^{\prime }\left( t,T_{1},T_{2}\right) \right]\tau_{d}\left( T_{1},T_{2}\right) },
\end{equation}
\begin{eqnarray}
BA_{fd}^{\prime }\left( t,T_{1},T_{2}\right) &=&\frac{F_{f}\left(
t;T_{1},T_{2}\right) \tau_{f}\left( T_{1},T_{2}\right) -F_{d}\left(
t;T_{1},T_{2}\right) \tau_{d}\left( T_{1},T_{2}\right) }{\tau_{d}\left(
T_{1},T_{2}\right) }  \notag \\
&=&\frac{1}{\tau_{d}\left( T_{1},T_{2}\right) }\left[ \frac{P_{f}\left(
t,T_{1}\right) }{P_{f}\left( t,T_{2}\right) }-\frac{P_{d}\left(
t,T_{1}\right) }{P_{d}\left( t,T_{2}\right) }\right]  \notag \\
&=&F_{d}\left( t;T_{1},T_{2}\right) \left[ BA_{fd}\left(
t,T_{1},T_{2}\right) -1\right] ,  \label{FwdBasisAdj3}
\end{eqnarray}%
which is more useful for comparisons with the market basis spreads of fig. %
\ref{FigBasisSwapReuters}. Notice that if $\complement_{d}=\complement_{f}$
we recover the single-curve case $BA_{fd}\left( t,T_{1},T_{2}\right) =1$,
$BA_{fd}^{\prime }\left( t,T_{1},T_{2}\right) =0.$

We stress that the forward basis in eqs. \ref{FwdBasisAdj2}-\ref{FwdBasisAdj3} is a straightforward consequence of the assumptions above,
essentially the existence of two yield curves and no arbitrage. Its
advantage is that it allows for a direct computation of the forward basis
between \emph{forward rates} for any time interval $\left[ T_{1},T_{2}%
\right] $, which is the relevant quantity for pricing and hedging interest
rate derivatives. In practice its value depends on the market basis spread
between the quotations of the two sets of vanilla instruments used in the
bootstrapping of the two curves $\complement_{d}$ and $\complement_{f}$.
On the other side, the limit of expressions \ref{FwdBasisAdj2}-\ref{FwdBasisAdj3} is that they reflect the \emph{statical}\footnote{we remind that the discount factors in eqs. \ref{FwdBasisAdj2}-\ref{FwdBasisAdj2} are calculated on the curves $\complement_{d}$, $\complement_{f}$ following the recipe described in sec. \ref{SecMultipleCurve}, not
using any dynamical model for the evolution of the rates.} differences
between the two interest rate markets $M_{d}$, $M_{f}$ carried by the two
curves $\complement_{d}$, $\complement_{f}$, but they are completely
independent of the interest rate dynamics in $M_{d}$ and $M_{f}$.

Notice also that the approach can be inverted to bootstrap a new yield curve
from a given yield curve plus a given forward basis, using the following
recursive relations
\begin{eqnarray}
P_{d,i} &=&
\frac{P_{f,i}BA_{fd,i}}{P_{f,i-1}-P_{f,i}+P_{f,i}BA_{fd,i}}P_{d,i-1}  \notag \\
&=&\frac{P_{f,i}}{P_{f,i-1}-P_{f,i}BA_{fd,i}^{\prime }\tau _{d,i}}P_{d,i-1},
\\
P_{f,i} &=&\frac{P_{d,i}}{P_{d,i}+\left( P_{d,i-1}-P_{d,i}\right) BA_{fd,i}}%
P_{f,i-1}  \notag \\
&=&\frac{P_{d,i}}{P_{d,i}+P_{d,i-1}BA_{fd,i}^{\prime }\tau _{d,i}}P_{f,i-1},
\end{eqnarray}%
where we have inverted eqs. \ref{FwdBasisAdj2}, \ref{FwdBasisAdj3} and
shortened the notation by putting $\tau_{x}\left( T_{i-1},T_{i}\right)
:=\tau_{x,i}$, $P_{x}\left( t,T_{i}\right) :=P_{x,i}$,
$BA_{fd}\left(t,T_{i-1},T_{i}\right) := BA_{fd,i}$.
Given the yield curve $x$ up to step $P_{x,i-1}$ plus the forward basis for the step $i-1\rightarrow i$, the equations above can be used to obtain the next step $P_{x,i}$.

\begin{figure}[tb]
\centering
\includegraphics[scale=1.0]{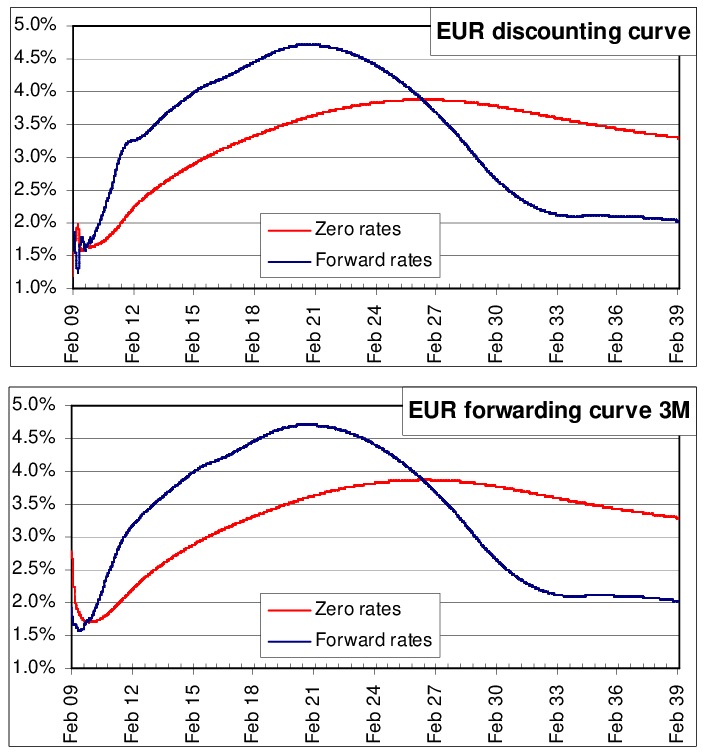}
%\vspace{-0.4cm}
\caption{EUR discounting curve $\complement_{d}$
(upper panel) and 3M forwarding curve $\complement_{3M}$ (lower panel)
at end of day Feb. 16th 2009. Blue lines: 3M-tenor forward rates $F\left(
t_{0};t,t+3M,\text{\emph{act/360}}\right) $, $t$ daily sampled and spot
date $t_{0}=$ 18th Feb. 2009; red lines: zero rates $F\left( t_{0};t,\text{\emph{act/365}}\right) $. Similar patterns are observed also in the 1M,
6M, 12M curves (not shown here, see ref. \protect\cite{AmeBia09}).}
\label{FigYieldCurves}
\end{figure}

\begin{figure}[htb]
\centering
\includegraphics[scale=0.6]{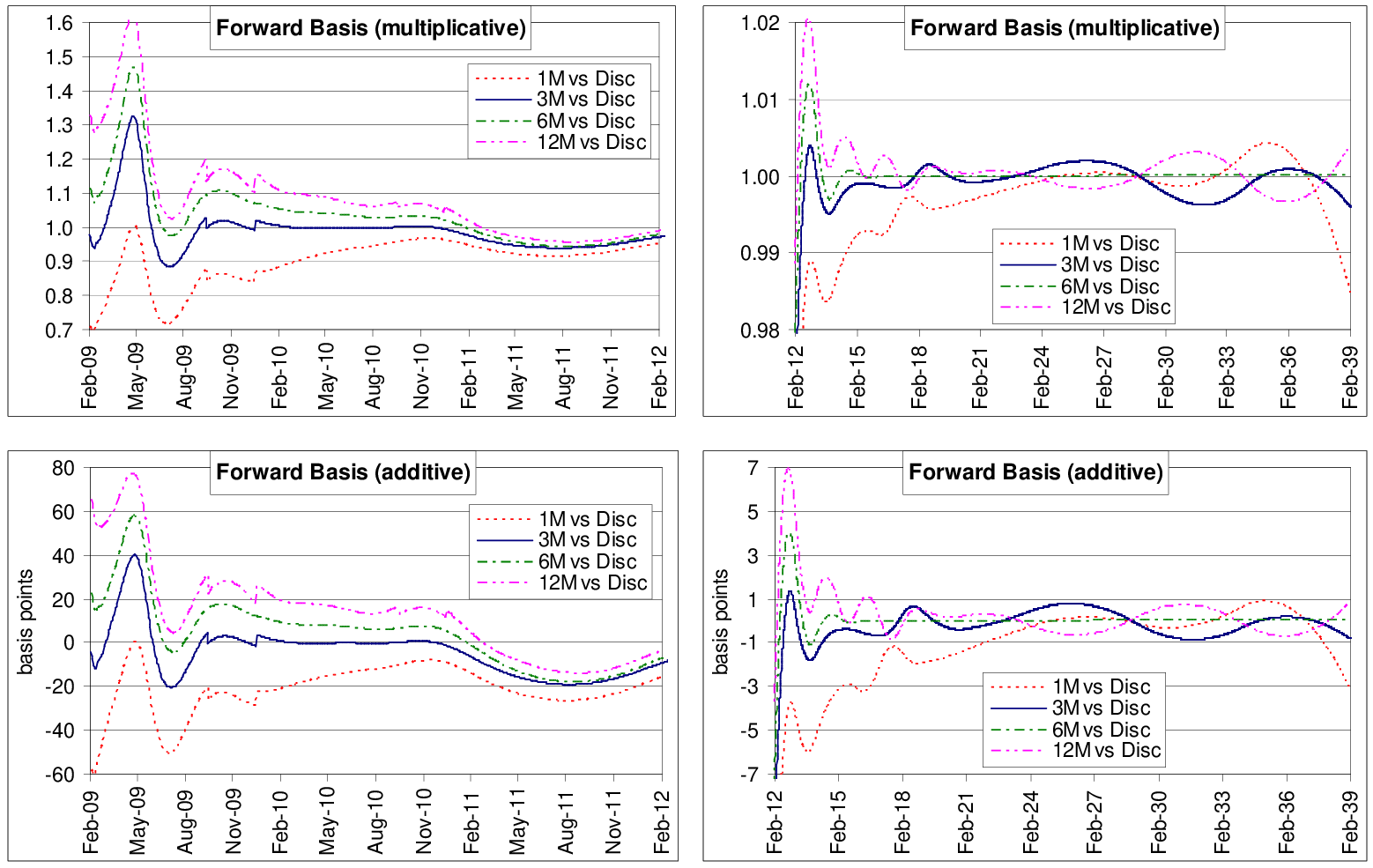}
%\vspace{-0.4cm}
\caption{upper panels: multiplicative basis adjustments from eq. \protect\ref{FwdBasisAdj2} as of end of day Feb. 16th, 2009, for daily sampled 3M-tenor forward rates as in fig. \protect\ref{FigYieldCurves}, calculated on $\complement_{1M}$, $\complement_{3M}$, $\complement_{6M}$ and $\complement_{12M}$ curves against $\complement_{d}$ taken as reference curve. Lower panels: equivalent plots of the additive basis adjustment of eq. \protect\ref{FwdBasisAdj3} between the same forward rates (basis points). Left panels: 0Y-3Y data; Right panels: 3Y-30Y data on magnified scales. The higher short-term adjustments seen in the left panels are due to the higher
short-term market basis spread (see Figs. \protect\ref{FigBasisSwapReuters}). The oscillating term structure observed is due to the amplification of small differences in the term structures of the curves.}
\label{FigBasisAdj}
\end{figure}

\begin{figure}[tb]
\centering
\includegraphics[scale=0.75]{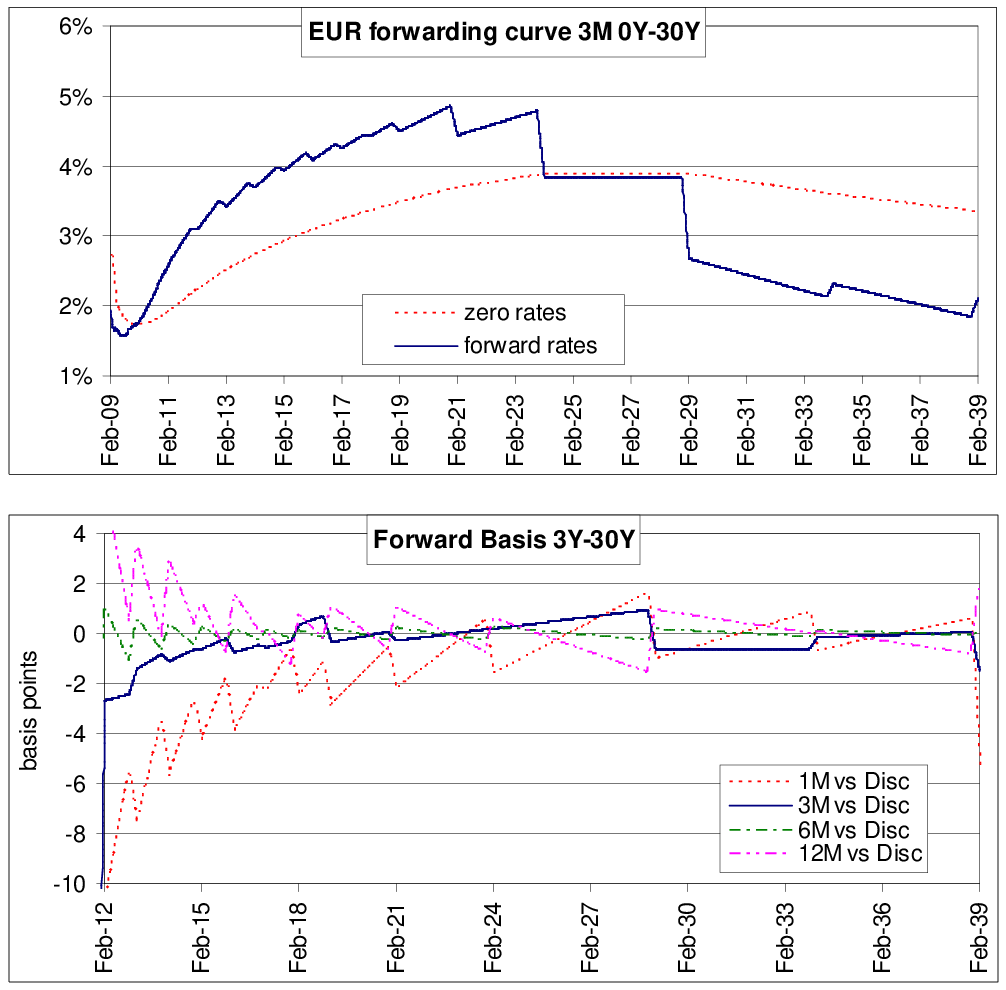}
%\vspace{-0.4cm}
\caption{the effect of poor interpolation schemes
(linear on zero rates, a common choice, see ref. \protect\cite{AmeBia09}) on
zero rates (upper panel, red line) 3M forward rates (upper panel, blue line)
and basis adjustments (lower panel). While the zero curve looks smooth, the
sag-saw shape of the forward curve clearly show the inadequacy of the
bootstrap, and the oscillations in the basis adjustment allow to further
appreciate the artificial differences induced in similar instruments priced
on the two curves.}
\label{FigBasisAdjLinearInterp}
\end{figure}

We now discuss a numerical example of the forward basis in a realistic
market situation. We consider the four interest rate underlyings $\mathbf{I}%
=\{I_{1M}$, $I_{3M}$, $I_{6M}$, $I_{12M}\}$, where $I\ =$ Euribor index, and
we bootstrap from market data five distinct yield curves $\mathbf{%
\complement }=\{\complement_{d}$, $\complement_{1M}$, $\complement_{3M}$,
$\complement_{6M}$, $\complement_{12M}\},$ using the first one for
discounting and the others for forwarding. We follow the methodology
described in ref. \cite{AmeBia09} using the corresponding open-source
development available in the QuantLib framework \cite{QuantLib}. The
discounting curve $\complement_{d}$ is built following a \textquotedblleft
pre-crisis\textquotedblright\ traditional recipe from the most liquid
deposit, IMM\ Futures/FRA on Euribor3M and swaps on Euribor6M. The other
four forwarding curves are built from convenient selections of depos, FRAs,
Futures, swaps and basis swaps with homogeneous underlying rate tenors; a
smooth and robust algorithm (monotonic cubic spline on log discounts) is
used for interpolations. Different choices (e.g. an Eonia discounting curve)
as well as other technicalities of the bootstrapping described in ref. \cite%
{AmeBia09} obviously would lead to slightly different numerical results, but
do not alter the conclusions drawn here.

In fig. \ref{FigYieldCurves} we plot both the 3M-tenor forward rates and the
zero rates calculated on $\complement_{d}$ and $\complement_{3M}$ as of
16th Feb. 2009 cob\footnote{%
close of business.}. Similar patterns are observed also in the other 1M, 6M,
12M curves (not shown here, see ref. \cite{AmeBia09}). In fig. \ref{FigBasisAdj} (upper panels) we plot the term structure of the four
corresponding multiplicative forward basis curves $\complement_{f}-\complement_{d}$ calculated through eq. \ref{FwdBasisAdj2}. In the
lower panels we also plot the additive forward basis given by eq. \ref{FwdBasisAdj3}. We observe in particular that the higher short-term basis
adjustments (left panels) are due to the higher short-term market basis
spreads (see fig. \ref{FigBasisSwapReuters}). Furthermore, the
medium-long-term $\complement_{6M}-\complement_{d}$ basis (dash-dotted
green lines in the right panels) are close to 1 and 0, respectively, as
expected from the common use of 6M swaps in the two curves. A similar, but
less evident, behavior is found in the short-term $\complement_{3M}-\complement_{d}$ basis (continuous blue line in the left panels), as
expected from the common 3M\ Futures and the uncommon deposits. The two
remaining basis curves $\complement_{1M}-\complement_{d}$ and $\complement_{12M}-\complement_{d}$ are generally far from 1 or 0 because of different
bootstrapping instruments. Obviously such details depend on our arbitrary
choice of \ the discounting curve.

Overall, we notice that all the basis curves $\complement_{f}-\complement_{d}$ reveal a complex micro-term structure, not present either in the
monotonic basis swaps market quotes of fig. \ref{FigBasisSwapReuters} or in
the smooth yield curves $\complement_{x}$. Such effect is essentially due
to an amplification mechanism of small local differences between the $%
\complement_{d}$ and $\complement_{f}$ forward curves. In fig. \ref{FigBasisAdjLinearInterp} we also show that smooth yield curves are a
crucial input for the forward basis: using a non-smooth bootstrapping
(linear interpolation on zero rates, still a diffused market practice), the
zero curve apparently shows no particular problems, while the forward curve
displays a sagsaw shape inducing, in turn, strong and unnatural oscillations
in the forward basis.

We conclude that, once a smooth and robust bootstrapping technique for yield
curve construction is used, the richer term structure of the forward basis
curves provides a sensitive indicator of the tiny, but observable, statical
differences between different interest rate market sub-areas in the post
credit crunch interest rate world, and a tool to assess the degree of
liquidity and credit issues in interest rate derivatives' prices. It is also
helpful for a better explanation of the profit\&loss encountered when
switching between the single- and the multiple-curve worlds.

\section{\label{SecQuantoAdj}Foreign-Currency Analogy and Quanto Adjustment}
A second important issue regarding no-arbitrage arises in the multiple-curve
framework. From eq. \ref{Cashflow} we have that, for instance, the
single-curve FRA price in eq. \ref{FRAPrice} is generalised into the
following multiple-curve expression%
\begin{equation}
\text{\textbf{FRA}}\left( t;T_{1},T_{2},K,N\right) =NP_{d}\left(
t,T_{2}\right) \tau_{f}\left( T_{1},T_{2}\right) \left\{ \mathbb{E}%
_{t}^{Q_{d}^{T_{2}}}\left[ L_{f}\left( T_{1},T_{2}\right) \right] -K\right\}
.  \label{FRAPrice1}
\end{equation}%
Instead, the current market practice is to price such FRA simply as
\begin{equation}
\text{\textbf{FRA}}\left( t;T_{1},T_{2},K,N\right) \simeq NP_{d}\left(
t,T_{2}\right) \tau_{f}\left( T_{1},T_{2}\right) \left[ F_{f}\left(
t;T_{1},T_{2}\right) -K\right] .  \label{FRAPriceWrong}
\end{equation}%
Obviously the forward rate $F_{f}\left( t;T_{1},T_{2}\right) $ is not, in
general, a martingale under the discounting measure $Q_{d}^{T_{2}}$, thus eq.
\ref{FRAPriceWrong} is just an approximation that discards the adjustment coming from this measure mismatch.
Hence, a theoretically correct pricing within the multiple-curve
framework requires the computation of expectations as in eq. \ref{FRAPrice1}
above. This will involve the \emph{dynamic} properties of the two interest
rate markets $M_{d}$ and $M_{f}$, or, in other words, it will require to
\emph{model} the dynamics for the interest rates in $M_{d}$ and $M_{f}$.
This task is easily accomplished by resorting to the natural analogy with
cross-currency derivatives. Going back to the beginning of sec. \ref{SecNotation}, we can identify $M_{d}$ and $M_{f}$ with the \emph{domestic}
and \emph{foreign} markets, $\complement_{d}$ and $\complement_{f}$ with
the corresponding curves, and the bank accounts $B_{d}\left( t\right) $, $B_{f}\left( t\right) $ with the corresponding currencies, respectively
\footnote{notice the lucky notation used, where \textquotedblleft \emph{d}%
\textquotedblright\ stands either for \textquotedblleft \emph{discounting}%
\textquotedblright\ or\textquotedblleft \emph{domestic}\textquotedblright\
and \textquotedblleft \emph{f}\textquotedblright\ for \textquotedblleft
\emph{forwarding}\textquotedblright\ or \textquotedblleft \emph{foreign}%
\textquotedblright , respectively.}. Within this framework, we can recognize
on the r.h.s of eq. \ref{NoArbitrage2} the forward discount factor from time
$T_{2}$ to time $T_{1}$ expressed in domestic currency, and on the r.h.s. of
eq. \ref{FRAPrice1} the expectation of the foreign forward rate w.r.t the
domestic forward measure. Hence, the computation of such expectation must
involve the quanto adjustment commonly encountered in the pricing of
cross-currency derivatives. The derivation of such adjustment can be found
in standard textbooks. Anyway, in order to fully appreciate the parallel
with the present double-curve-single-currency case, it is useful to run
through it once again. In particular, we will adapt to the present context
the discussion found in ref. \cite{BriMer06}, chs. 2.9 and 14.4.

\subsection{Forward Rates\label{SecQAFwdRates}}
\begin{figure}[tb]
\centering
\includegraphics[scale=0.75]{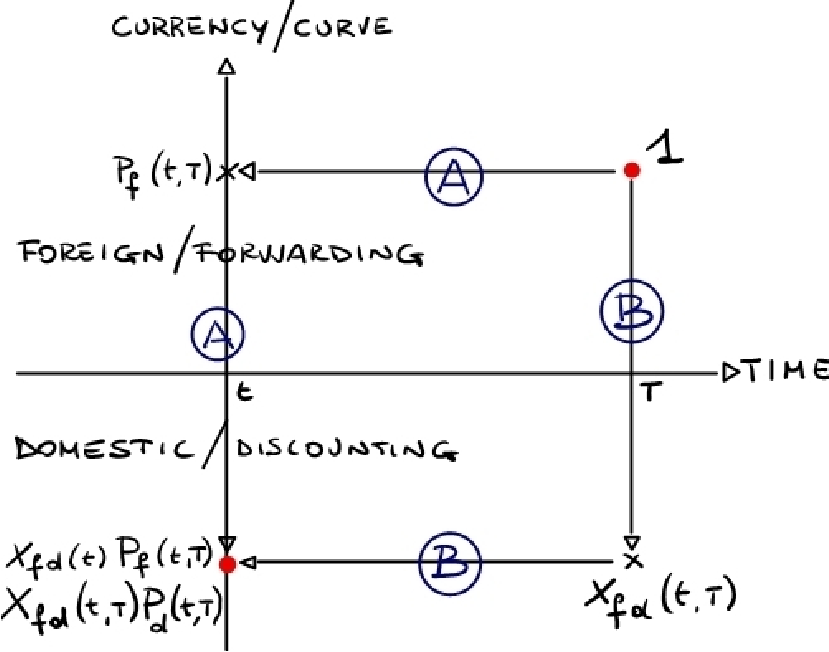}
%\vspace{-0.4cm}
\caption{Picture of
no-arbitrage interpretation for the forward exchange rate in eq. \protect\ref{FwdExchRate}. Moving, in the yield curve vs time plane, from top right to
bottom left corner through path A or path B must be equivalent.
Alternatively, we may think to no-arbitrage as a sort of zero
\textquotedblleft circuitation\textquotedblright , sum of all trading events
following a closed path starting and stopping at the same point in the
plane. This description is equivalent to the traditional \textquotedblleft
table of transaction\textquotedblright\ picture, as found e.g. in fig. 1 of
ref. \protect\cite{TucPor03}.}
\label{FigFwdExchRate}
\end{figure}

In the double--curve-double-currency case, no arbitrage requires the
existence at any time $t_{0}\leq t\leq T$ of a spot and a forward exchange
rate between equivalent amounts of money in the two currencies such that
\begin{gather}
c_{d}\left( t\right) =x_{fd}\left( t\right) c_{f}\left( t\right) ,
\label{SpotExchRate} \\
X_{fd}\left( t,T\right) P_{d}\left( t,T\right) =x_{fd}\left( t\right)
P_{f}\left( t,T\right) ,  \label{FwdExchRate}
\end{gather}%
where the subscripts \emph{f} and \emph{d} stand for \emph{foreign}
and \emph{domestic}, $c_{d}\left( t\right) $ is any cash flow (amount of
money) at time $t$ in units of domestic-currency and $c_{f}\left( t\right) $
is the corresponding cash flow at time $t$ (the corresponding amount of
money) in units of foreign currency. Obviously $X_{fd}\left( t,T\right)
\rightarrow x_{fd}\left( t\right) $ for $t\rightarrow T$. Expression \ref{FwdExchRate} is still a consequence of no arbitrage. This can be understood
with the aid of fig. \ref{FigFwdExchRate}: starting from top right corner in
the time vs currency/yield curve plane with an unitary cash flow at time $T>t$
in foreign currency, we can either move along path A by discounting at time $%
t$ on curve $\complement_{f}$ using $P_{f}\left( t,T\right) $ and then by
changing into domestic currency units using the spot exchange rate $%
x_{fd}\left( t\right) $, ending up with $x_{fd}\left( t\right) P_{f}\left(
t,T\right) $ units of domestic currency; or, alternatively, we can follow
path B by changing at time $T$ into domestic currency units using the
forward exchange rate $X_{fd}\left( t,T\right) $ and then by discounting on $%
\complement_{d}$ using $P_{d}\left( t,T\right) $, ending up with $X_{fd}\left( t,T\right) P_{d}\left( t,T\right) $ units of domestic currency.
Both paths stop at bottom left corner, hence eq. \ref{FwdExchRate} must hold
by no arbitrage.

Now, our double-curve-single-currency case is immediately obtained from the
discussion above by thinking to the subscripts \emph{f} and \emph{d} as
shorthands for \emph{forwarding} and \emph{discounting} and by
recognizing that, having a single currency, the \emph{today's} spot exchange rate must
collapse to 1, $x_{fd}\left( t_0\right) = 1$.
Obviously for $\complement_{d}=\complement_{f}$ we recover the
single-currency, single-curve case $X_{fd}\left( t,T\right) =1$ $\forall $ $t,T$.
The financial interpretation of the forward exchange rate in eq. \ref{FwdExchRate} within this framework is straightforward: it is nothing else
that the counterparty of the forward\ basis in eq. \ref{FwdBasisAdj} for
discount factors on the two yield curves $\complement_{d}$ and $\complement_{f}$.
Substituting eq. \ref{FwdExchRate} into eq. \ref{FwdBasisAdj} we obtain the following relation
\begin{equation}
BA_{fd}\left( t,T_{1},T_{2}\right)
= \frac{P_d\left(t,T_1\right)X_{fd}\left(t,T_1\right) - P_d\left(t,T_2\right)X_{fd}\left(t,T_2\right)}
{X_{fd}(t,T_2)\left[P_d\left(t,T_1\right) - P_{d}\left(t,T_{2}\right)\right]}.
\end{equation}
\par
We proceed by assuming, according to the standard market practice, the
following (driftless) lognormal martingale dynamic for $\complement_{f}$
(foreign) forward rates%
\begin{equation}
\frac{dF_{f}\left( t;T_{1},T_{2}\right) }{F_{f}\left( t;T_{1},T_{2}\right) }%
=\sigma _{f}\left( t\right) dW_{f}^{T_{2}}\left( t\right) ,\;\;t\leq T_{1},
\label{FwdDynamic}
\end{equation}%
where $\sigma _{f}\left( t\right) $ is the volatility (positive
deterministic function of time) of the process, under the probability space $%
\left( \Omega ,\mathcal{F}^{f},Q_{f}^{T_{2}}\right) $ with the filtration $%
\mathcal{F}_{t}^{f}$ generated by the brownian motion $W_{f}^{T_{2}}$ under
the forwarding (foreign) $T_{2}-$forward measure $Q_{f}^{T_{2}}$, associated
to the $\complement_{f}$ (foreign) numeraire $P_{f}\left( t,T_{2}\right) $.

Next, since $X_{fd}\left( t,T_{2}\right) $ in eq. \ref{FwdExchRate} is the
ratio between the price at time $t$ of a $\complement_{d}$ (domestic)
tradable asset $x_{fd}\left(t\right)P_{f}\left(t,T_{2}\right)$ and the $\complement_{d}$ numeraire $P_{d}\left(t,T_{2}\right) $, it must evolve according to a (driftless)\
martingale process under the associated discounting (domestic) $T_{2}-$
forward measure $Q_{d}^{T_{2}}$,
\begin{equation}
\frac{dX_{fd}\left( t,T_{2}\right) }{X_{fd}\left( t,T_{2}\right) }=\sigma
_{X}\left( t\right) dW_{X}^{T_{2}}\left( t\right) ,\;\;t\leq T_{2},
\end{equation}%
where $\sigma _{X}\left( t\right) $ is the\ volatility (positive
deterministic function of time) of the process and $W_{X}^{T_{2}}$ is a
brownian motion under $Q_{d}^{T_{2}}$ such that%
\begin{equation}
dW_{f}^{T_{2}}\left( t\right) dW_{X}^{T_{2}}\left( t\right) =\rho
_{fX}\left( t\right) dt.
\end{equation}

Now, in order to calculate expectations such as in the r.h.s. of eq. \ref{FRAPrice1}, we must switch from the forwarding (foreign) measure $%
Q_{f}^{T_{2}}$ associated to the numeraire $P_{f}\left( t,T_{2}\right) $ to
the discounting (domestic) measure $Q_{d}^{T_{2}}$ associated to the
numeraire $P_{d}\left( t,T_{2}\right) $. In our double-curve-single-currency
language this amounts to transform a cash flow on curve $\complement_{f}$ to
the corresponding cash flow on curve $\complement_{d}$. Recurring to the
change-of-numeraire technique (see refs. \cite{BriMer06}, \cite{Jam1989},
\cite{GER1995}) we obtain that the dynamic of $F_{f}\left(
t;T_{1},T_{2}\right) $ under $Q_{d}^{T_{2}}$ acquires a non-zero drift%
\begin{gather}
\frac{dF_{f}\left( t;T_{1},T_{2}\right) }{F_{f}\left( t;T_{1},T_{2}\right) }%
=\mu _{f}\left( t\right) dt+\sigma _{f}\left( t\right) dW_{f}^{T_{2}}\left(
t\right) ,\;\;t\leq T_{1}, \\
\mu _{f}\left( t\right) =-\sigma _{f}\left( t\right) \sigma _{X}\left(
t\right) \rho _{fX}\left( t\right) ,
\label{DriftAdjFwd}
\end{gather}%
and that $F_{f}\left( T_{1};T_{1},T_{2}\right) $ is lognormally distributed
under $Q_{d}^{T_{2}}$ with mean and variance given by
\begin{gather}
\mathbb{E}_{t}^{Q_{d}^{T_{2}}}\left[ \ln \frac{F_{f}\left(
T_{1};T_{1},T_{2}\right) }{F_{f}\left( t;T_{1},T_{2}\right) }\right]
=\int_{t}^{T_{1}}\left[ \mu _{f}\left( u\right) -\frac{1}{2}\sigma
_{f}^{2}\left( u\right) \right] du, \\
\text{Var}_{t}^{Q_{d}^{T_{2}}}\left[ \ln \frac{F_{f}\left(
T_{1};T_{1},T_{2}\right) }{F_{f}\left( t;T_{1},T_{2}\right) }\right]
=\int_{t}^{T_{1}}\sigma _{f}^{2}\left( u\right) du.
\end{gather}%
We thus obtain the following expressions, for $t_{0}\leq t<T_{1}$,%
\begin{align}
\mathbb{E}_{t}^{Q_{d}^{T_{2}}}\left[ F_{f}\left( T_{1};T_{1},T_{2}\right) %
\right] & =F_{f}\left( t;T_{1},T_{2}\right) QA_{fd}\left( t,T_{1},\sigma
_{f},\sigma _{X},\rho _{fX}\right) ,  \label{QuantoAdj} \\
QA_{fd}\left( t,T_{1},\sigma _{f},\sigma _{X},\rho _{fX}\right) & =\exp
\int_{t}^{T_{1}}\mu _{f}\left( u\right) du  \notag \\
& =\exp \left[ -\int_{t}^{T_{1}}\sigma _{f}\left( u\right) \sigma _{X}\left(
u\right) \rho _{fX}\left( u\right) du\right] ,  \label{QuantoExp}
\end{align}%
where $QA_{fd}\left( t,T_{1},\sigma _{f},\sigma _{X},\rho _{fX}\right) $ is
the (multiplicative)\ quanto adjustment. We may also define an additive
quanto adjustment as%
\begin{gather}
\mathbb{E}_{t}^{Q_{d}^{T_{2}}}\left[ F_{f}\left( T_{1};T_{1},T_{2}\right) %
\right] =F_{f}\left( t;T_{1},T_{2}\right) +QA_{fd}^{\prime }\left(
t,T_{1},\sigma _{f},\sigma _{X},\rho _{fX}\right) ,  \label{QuantoAdj2} \\
QA_{fd}^{\prime }\left( t,T_{1},\sigma _{f},\sigma _{X},\rho _{fX}\right)
=F_{f}\left( t;T_{1},T_{2}\right) \left[ QA_{fd}\left( t,T_{1},\sigma
_{f},\sigma _{X},\rho _{fX}\right) -1\right] ,  \label{QuantoAdd}
\end{gather}%
where the second relation comes from eq. \ref{QuantoAdj}. Finally, combining
eqs. \ref{QuantoAdj}, \ref{QuantoAdj2} with eqs. \ref{FwdBasisAdj2}, \ref{FwdBasisAdj3} we may derive a relation between the quanto and the basis adjustments,
\begin{equation}
BA_{fd}\left(t,T_{1},T_{2}\right)QA_{fd}\left( t,T_{1},\sigma_{f},\sigma _{X},\rho _{fX}\right) =\frac{\tau_f\left(T_{1},T_{2}\right)\mathbb{E}_{t}^{Q_{d}^{T_{2}}}\left[ L_{f}\left( T_{1},T_{2}\right) \right] }
{\tau_d\left(T_{1},T_{2}\right)\mathbb{E}_{t}^{Q_{d}^{T_{2}}}\left[ L_{d}\left( T_{1},T_{2}\right) \right]},
\end{equation}
\begin{multline}
BA_{fd}^{\prime }\left(t,T_{1},T_{2}\right)\tau_d\left(T_{1},T_{2}\right) +
QA_{fd}^{\prime }\left(t,T_{1},\sigma _{f},\sigma _{X},\rho _{fX}\right)\tau_f\left(T_{1},T_{2}\right) \\
= \mathbb{E}_{t}^{Q_{d}^{T_{2}}}\left[L_{f}\left(T_{1},T_{2}\right) \right]\tau_f\left(T_{1},T_{2}\right)
- \mathbb{E}_{t}^{Q_{d}^{T_{2}}}\left[L_{d}\left(T_{1},T_{2}\right) \right]\tau_d\left(T_{1},T_{2}\right)
\end{multline}
for multiplicative and additive adjustments, respectively.
\par
We conclude that the foreign-currency analogy allows us to compute the
expectation in eq. \ref{FRAPrice1} of a forward rate on curve $\complement_{f}$ \ w.r.t. the discounting measure $Q_{d}^{T_{2}}$ in terms of a
well-known quanto adjustment, typical of cross-currency derivatives. Such
adjustment naturally follows from a change between the\ $T$-forward
probability measures $Q_{f}^{T_{2}}$ and $Q_{d}^{T_{2}}$, or numeraires $%
P_{f}\left( t,T_{2}\right) $ and $P_{d}\left( t,T_{2}\right) $, associated
to the two yield curves, $\complement_{f}$ and $\complement_{d},$
respectively. Notice that the expression \ref{QuantoExp} depends on the
average over the time interval $\left[ t,T_{1}\right] $ of the product of
the volatility $\sigma _{f}$ of the $\complement_{f}$ (foreign) forward
rates $F_{f}$, of the volatility $\sigma _{X}$ of the forward exchange rate $X_{fd}$ between curves $\complement_{f}$ and $\complement_{d}$, and of the
correlation $\rho _{fX}$ between $F_{f}$ and $X_{fd}$. It does not depend
either on the volatility $\sigma _{d}$ of the $\complement_{d}$ (domestic)\
forward rates $F_{d}$ or on any stochastic quantity after time $T_{1}$. The
latter fact is actually quite natural, because the stochasticity of the
forward rates involved ceases at their fixing time $T_{1}$\emph{. }The
dependence on the cash flow time $T_{2}$ is actually implicit in eq \ref{QuantoExp}, because the volatilities and the correlation involved are
exactly those of the forward and exchange rates on the time interval $\left[
T_{1},T_{2}\right] $. Notice in particular that a non-trivial adjustment is
obtained if and only if the forward exchange rate $X_{fd}$ is \emph{%
stochastic} ($\sigma _{X}\neq 0$) and \emph{correlated} to the forward
rate $F_{f}$ ($\rho _{fX}\neq 0$); otherwise expression \ref{QuantoExp}
collapses to the single curve case $QA_{fd}=1$.

\subsection{Swap Rates\label{SecQASwapRates}}
The discussion above can be remapped, with some attention, to swap rates.
Given two increasing dates sets $\mathbf{T=}\left\{ T_{0},...,T_{n}\right\}$,
$\mathbf{S=}\left\{ S_{0},...,S_{m}\right\} $, $T_{0}=S_{0}\geq t$ and an
interest rate swap with a floating leg paying at times $T_{i}$, $i=1,..,n$,
the Xibor rate with tenor $\left[ T_{i-1},T_{i}\right] $ fixed at time $%
T_{i-1}$, plus a fixed leg paying at times $S_{j}$, $j=1,..,m$, a fixed
rate, the corresponding fair swap rate $S_{f}\left( t,\mathbf{T},\mathbf{S}%
\right) $ on curve $\complement_{f}$ is defined by the following
equilibrium (no arbitrage) relation between the present values of the two
legs,%
\begin{equation}
S_{f}\left( t,\mathbf{T},\mathbf{S}\right) A_{f}\left( t,\mathbf{S}\right)
=\sum\limits_{i=1}^{n}P_{f}\left( t,T_{i}\right) \tau_{f}\left(
T_{i-1},T_{i}\right) F_{f}\left( t;T_{i-1},T_{i}\right) ,\;\;t\leq
T_{0}=S_{0},  \label{SwapRateFwd}
\end{equation}%
where
\begin{equation}
A_{f}\left( t,\mathbf{S}\right) =\sum\limits_{j=1}^{m}P_{f}\left(
t,S_{j}\right) \tau_{f}\left( S_{j-1},S_{j}\right)
\end{equation}%
is the annuity on curve $\complement_{f}$. Following the standard market
practice, we observe that, assuming the annuity as the numeraire on curve $\complement_{f}$, the swap rate in eq. \ref{SwapRateFwd} is the ratio between a tradable asset (the value of the swap floating leg on curve $\complement_{f}$)\ and the numeraire $A_{f}\left( t,\mathbf{S}\right) $,
and thus it is a martingale under the associated forwarding (foreign) swap
measure $Q_{f}^{\mathbf{S}}$. Hence we can assume, as in eq. \ref{FwdDynamic}, a driftless geometric brownian motion for the swap rate under $Q_{f}^{\mathbf{S}}$,
\begin{equation}
\frac{dS_{f}\left( t,\mathbf{T},\mathbf{S}\right) }{S_{f}\left( t,\mathbf{T},%
\mathbf{S}\right) }=\nu _{f}\left( t,\mathbf{T},\mathbf{S}\right) dW_{f}^{%
\mathbf{T},\mathbf{S}}\left( t\right) ,\;\;t\leq T_{0},
\end{equation}%
where $\upsilon _{f}\left( t,\mathbf{T},\mathbf{S}\right) $ is the
volatility (positive deterministic function of time) of the process and $W_{f}^{\mathbf{T},\mathbf{S}}$ is a brownian motion under $Q_{f}^{\mathbf{S}}$. Then, mimicking the discussion leading to eq. \ref{FwdExchRate}, the following relation
\begin{eqnarray}
\sum\limits_{j=1}^{m}P_{d}\left( t,S_{j}\right) \tau_{d}\left(
S_{j-1},S_{j}\right) X_{fd}\left( t,S_{j}\right) &=&x_{fd}\left( t\right)
\sum\limits_{j=1}^{m}P_{f}\left( t,S_{j}\right) \tau_{f}\left(
S_{j-1},S_{j}\right)  \notag \\
&=&x_{fd}\left(t\right)A_{f}\left( t,\mathbf{S}\right)
\end{eqnarray}%
must hold by no arbitrage between the two curves $\complement_{f}$ and $\complement_{d}$. Defining a swap forward exchange rate $Y_{fd}\left( t,\mathbf{S}\right) $ such that
\begin{eqnarray}
x_{fd}\left(t\right) A_{f}\left( t,\mathbf{S}\right) , &=&\sum\limits_{j=1}^{m}P_{d}\left(
t,S_{j}\right) \tau_{d}\left( S_{j-1},S_{j}\right) X_{fd}\left(
t,S_{j}\right)  \notag \\
&=&Y_{fd}\left( t,\mathbf{S}\right) \sum\limits_{j=1}^{m}P_{d}\left(
t,S_{j}\right) \tau_{d}\left( S_{j-1},S_{j}\right) =Y_{fd}\left( t,\mathbf{S%
}\right) A_{d}\left( t,\mathbf{S}\right) ,
\end{eqnarray}%
we obtain the expression%
\begin{equation}
Y_{fd}\left( t,\mathbf{S}\right)
= x_{fd}\left(t\right) \frac{A_{f}\left(t,\mathbf{S}\right)}{A_{d}\left(t,\mathbf{S}\right)},
\end{equation}%
equivalent to eq. \ref{FwdExchRate}. Hence, since $Y_{fd}\left( t,\mathbf{S}%
\right) $ is the ratio between the price at time $t$ of the $\complement_{d} $ (domestic) tradable asset $x_{fd}\left( t\right) A_{f}\left( t,%
\mathbf{S}\right) $ and the numeraire $A_{d}\left( t,\mathbf{S}\right) $, it
must evolve according to a (driftless)\ martingale process under the
associated discounting (domestic) swap measure $Q_{d}^{\mathbf{S}}$,%
\begin{equation}
\frac{dY_{fd}\left( t,\mathbf{S}\right) }{Y_{fd}\left( t,\mathbf{S}\right) }%
=\nu _{Y}\left( t,\mathbf{S}\right) dW_{Y}^{\mathbf{S}}\left( t\right)
,\;\;t\leq T_{0},
\end{equation}%
where $v_{Y}\left( t,\mathbf{S}\right) $ is the\ volatility (positive
deterministic function of time) of the process and $W_{Y}^{\mathbf{S}}$ is a
brownian motion under $Q_{d}^{\mathbf{S}}$ such that%
\begin{equation}
dW_{f}^{\mathbf{T},\mathbf{S}}\left( t\right) dW_{Y}^{\mathbf{S}}\left(
t\right) =\rho _{fY}\left( t,\mathbf{T},\mathbf{S}\right) dt.
\end{equation}

Now, applying again the change-of-numeraire technique of sec. \ref{SecQAFwdRates}, we obtain that the dynamic of the swap rate $S_{f}\left( t,%
\mathbf{T},\mathbf{S}\right) $ under the discounting (domestic) swap measure
$Q_{d}^{\mathbf{S}}$ acquires a non-zero drift%
\begin{gather}
\frac{dS_{f}\left( t,\mathbf{T},\mathbf{S}\right) }{S_{f}\left( t,\mathbf{T},%
\mathbf{S}\right) }=\lambda _{f}\left( t,\mathbf{T},\mathbf{S}\right) dt+\nu
_{f}\left( t,\mathbf{T},\mathbf{S}\right) dW_{f}^{\mathbf{T},\mathbf{S}%
}\left( t\right) ,\;\;t\leq T_{0}, \\
\lambda _{f}\left( t,\mathbf{T},\mathbf{S}\right) =-\nu _{f}\left( t,\mathbf{%
T},\mathbf{S}\right) \nu _{Y}\left( t,\mathbf{S}\right) \rho _{fY}\left( t,%
\mathbf{T},\mathbf{S}\right) ,
\label{DriftAdjSwap}
\end{gather}%
and that $S_{f}\left( t,\mathbf{T},\mathbf{S}\right) $ is lognormally
distributed under $Q_{d}^{\mathbf{S}}$ with mean and variance given by
\begin{eqnarray}
\mathbb{E}_{t}^{Q_{d}^{\mathbf{S}}}\left[ \ln \frac{S_{f}\left( T_{0},%
\mathbf{T},\mathbf{S}\right) }{S_{f}\left( t,\mathbf{T},\mathbf{S}\right) }%
\right] &=&\int_{t}^{T_{0}}\left[ \lambda _{f}\left( u,\mathbf{T},\mathbf{S}%
\right) -\frac{1}{2}\nu _{f}^{2}\left( u,\mathbf{T},\mathbf{S}\right) \right]
du, \\
\text{Var}_{t}^{Q_{d}^{\mathbf{S}}}\left[ \ln \frac{S_{f}\left( T_{0},%
\mathbf{T},\mathbf{S}\right) }{S_{f}\left( t,\mathbf{T},\mathbf{S}\right) }%
\right] &=&\int_{tf}^{T_{0}}\nu _{f}^{2}\left( u,\mathbf{T},\mathbf{S}%
\right) du.
\end{eqnarray}%
We thus obtain the following expressions, for $t_{0}\leq t<T_{0}$,%
\begin{align}
\mathbb{E}_{t}^{Q_{d}^{\mathbf{S}}}\left[ S_{f}\left( T_{0},\mathbf{T},%
\mathbf{S}\right) \right] & =S_{f}\left( t,\mathbf{T},\mathbf{S}\right)
QA_{fd}\left( t,\mathbf{T},\mathbf{S},\nu _{f},\nu _{Y},\rho _{fY}\right) ,
\label{QuantoAdjSwap} \\
QA_{fd}\left( t,\mathbf{T},\mathbf{S},\nu _{f},\nu _{Y},\rho _{fY}\right) &
=\exp \int_{t}^{T_{0}}\lambda _{f}\left( u,\mathbf{T},\mathbf{S}\right) du
\notag \\
& =\exp \left[ -\int_{t}^{T_{0}}\nu _{f}\left( u,\mathbf{T},\mathbf{S}%
\right) \nu _{Y}\left( u,\mathbf{S}\right) \rho _{fY}\left( u,\mathbf{T},%
\mathbf{S}\right) du\right]  \label{QuantoExpSwap}
\end{align}

The same considerations as in sec. \ref{SecQAFwdRates} apply. In particular,
we observe that the adjustment in eqs. \ref{QuantoAdjSwap}, \ref{QuantoAdjSwap2} naturally follows from a change between the\ probability
measures $Q_{f}^{\mathbf{S}}$ and $Q_{d}^{\mathbf{S}}$, or numeraires $%
A_{f}\left( t,\mathbf{S}\right) $ and $A_{d}\left( t,\mathbf{S}\right) $,
associated to the two yield curves, $\complement_{f}$ and $\complement_{d}, $ respectively, once swap rates are considered. In the EUR market, the
volatility $\nu _{f}\left( u,\mathbf{T},\mathbf{S}\right) $ in eq. \ref{QuantoExpSwap} can be extracted from quoted swaptions on Euribor6M, while
for other rate tenors and for $\nu _{Y}\left( u,\mathbf{S}\right) $ and $%
\rho _{fY}\left( u,\mathbf{T},\mathbf{S}\right) $ one must resort to
historical estimates.

An additive quanto adjustment can also be defined as before%
\begin{gather}
\mathbb{E}_{t}^{Q_{d}^{\mathbf{S}}}\left[ S_{f}\left( T_{0},\mathbf{T},%
\mathbf{S}\right) \right] =S_{f}\left( t,\mathbf{T},\mathbf{S}\right)
+QA_{fd}^{\prime }\left( t,\mathbf{T},\mathbf{S},\nu _{f},\nu _{Y},\rho
_{fY}\right) ,  \label{QuantoAdjSwap2} \\
QA_{fd}^{\prime }\left( t,\mathbf{T},\mathbf{S},\nu _{f},\nu _{Y},\rho
_{fY}\right) =S_{f}\left( t,\mathbf{T},\mathbf{S}\right) \left[
QA_{fd}\left( t,\mathbf{T},\mathbf{S},\nu _{f},\nu _{Y},\rho _{fY}\right) -1%
\right] .  \label{QuantoAddSwap}
\end{gather}

\section{\label{SecPricing}Double-Curve Pricing \&\ Hedging Interest Rate
Derivatives}

\subsection{Pricing}

The results of sec. \ref{SecQuantoAdj} above allows us to derive no
arbitrage, double-curve-single-currency pricing formulas for interest rate
derivatives. The recipes are, basically, eqs. \ref{QuantoAdj}-\ref{QuantoExp}
or \ref{QuantoAdjSwap}-\ref{QuantoExpSwap}.

The simplest interest rate derivative is a floating zero coupon bond paying
at time $T$ a single cash flow depending on a single spot rate (e.g. the
Xibor) fixed at time $t<T$,
\begin{equation}
\text{\textbf{ZCB}}\left( T;T,N\right) =N\tau_{f}\left( t,T\right)
L_{f}\left( t,T\right) .  \label{ZCBPayoff}
\end{equation}%
Being%
\begin{equation}
L_{f}\left( t,T\right) =\frac{1-P_{f}\left( t,T\right) }{\tau_{f}\left(
t,T\right) P_{f}\left( t,T\right) }=F_{f}\left( t;t,T\right) ,
\end{equation}%
the price at time $t\leq T$ is given by
\begin{eqnarray}
\text{\textbf{ZCB}}\left( t;T,N\right) &=&NP_{d}\left( t,T\right) \tau_{f}\left( t,T\right) \mathbb{E}_{t}^{Q_{d}^{T}}\left[ F_{f}\left(
t;t,T\right) \right]  \notag \\
&=&NP_{d}\left( t,T\right) \tau_{f}\left( t,T\right) L_{f}\left( t,T\right)
.  \label{ZCBprice}
\end{eqnarray}%
Notice that the forward basis in eq. \ref{ZCBprice} disappears and we are
left with the standard pricing formula, modified according to the
double-curve framework.

Next we have the FRA, whose payoff is given in eq. \ref{FRAPayoff} and whose
price at time $t\leq T_{1}$ is given by%
\begin{multline}
\text{\textbf{FRA}}\left( t;T_{1},T_{2},K,N\right)
= N P_{d}\left(t,T_{2}\right) \tau_{f}\left(T_{1},T_{2}\right)
\left\{\mathbb{E}_t^{Q_d^{T_2}}\left[F_f\left(T_1;T_1,T_2\right)\right]-K\right\} \\
= N P_d\left(t,T_2\right)\tau_f\left(T_1,T_2\right)
\left[F_f\left(t;T_1,T_2\right)QA_{fd}\left(t,T_1,\sigma_f,\sigma_X,\rho_{fX}\right) - K\right] .  \label{FRAPrice2}
\end{multline}
Notice that in eq. \ref{FRAPrice2} for $K=0$ and $T_{1}=t$ we recover the
zero coupon bond price in eq. \ref{ZCBprice}.
\par
For a (payer)\ floating vs fixed swap with payment dates vectors $\mathbf{T},\mathbf{S}$ as in sec. \ref{SecQASwapRates} we have the price at time $t\leq T_{0}$
\begin{align}
& \text{\textbf{Swap}}\left( t;\mathbf{T},\mathbf{S},\mathbf{K},\mathbf{N}\right)  \notag \\
& =\sum\limits_{i=1}^{n}N_{i}P_{d}\left( t,T_{i}\right)\tau_{f}\left(T_{i-1},T_{i}\right) F_f\left(t;T_{i-1},T_i\right) QA_{fd}\left(t,T_{i-1},\sigma_{f,i},\sigma_{X,i},\rho_{fX,i}\right)  \notag \\
& -\sum\limits_{j=1}^{m}N_{j}P_{d}\left( t,S_{j}\right) \tau_{d}\left(
S_{j-1},S_{j}\right) K_{j}.
\end{align}%
For constant nominal $N$ and fixed rate $K$ the fair (equilibrium) swap rate
is given by%
\begin{equation}
S_{f}\left( t,\mathbf{T},\mathbf{S}\right) =\frac{\sum\limits_{i=1}^{n}P_{d}%
\left( t,T_{i}\right) \tau_{f}\left( T_{i-1},T_{i}\right) F_{f}\left(
t;T_{i-1},T_{i}\right) QA_{fd}\left( t,T_{i-1},\sigma_{f,i},\sigma_{X,i},\rho _{fX,i}\right) }{A_{d}\left( t,\mathbf{S}\right) },
\label{SwapRate}
\end{equation}%
where
\begin{equation}
A_{d}\left( t,\mathbf{S}\right) =\sum\limits_{j=1}^{m}P_{d}\left(
t,S_{j}\right) \tau_{d}\left( S_{j-1},S_{j}\right)  \label{Annuity}
\end{equation}%
is the annuity on curve $\complement_{d}$.
\par
For caplet/floorlet options on a $T_{1}$-spot rate with payoff at maturity $%
T_{2}$ given by
\begin{equation}
\text{\textbf{cf}}\left( T_{2};T_{1},T_{2},K,\mathbf{\omega ,}N\right) =N%
\text{Max}\left\{ \mathbf{\omega }\left[ L_{f}\left( T_{1},T_{2}\right) -K%
\right] \right\} \tau_{f}\left( T_{1},T_{2}\right) ,
\end{equation}
the standard market-like pricing expression at time $t\leq T_{1}\leq T_{2}$
is modified as follows
\begin{multline}
\text{\textbf{cf}}\left( t;T_{1},T_{2},K,\mathbf{\omega ,}N\right)
= N \mathbb{E}_{t}^{Q_{d}^{T_{2}}}\left[ \text{Max}\left\{ \mathbf{\omega }\left[
L_{f}\left( T_{1},T_{2}\right) -K\right] \right\} \tau_{f}\left(T_{1},T_{2}\right) \right] \\
= N P_{d}\left( t,T_{2}\right) \tau_{f}\left( T_{1},T_{2}\right)
Bl\left[F_f\left(t;T_1,T_2\right)
QA_{fd}\left(t,T_1,\sigma_f,\sigma_X,\rho_{fX}\right),K,\mu_{f},\sigma_f,\omega \right],
\end{multline}
where $\omega =+/-1$ for caplets/floorlets, respectively, and
\begin{gather}
Bl\left[F,K,\mu ,\sigma ,\omega \right] =\omega \left[ F\Phi \left(\omega
d^{+}\right) -K\Phi \left( \omega d^{-}\right) \right] , \\
d^{\pm }=\frac{\ln \frac{F}{K}+\mu \left( t,T\right) \pm \frac{1}{2}\sigma
^{2}\left( t,T\right) }{\sigma \left( t,T\right) }, \\
\mu \left( t,T\right) =\int_{t}^{T}\mu \left( u\right) du,\;\;\;\sigma
^{2}\left( t,T\right) =\int_{t}^{T}\sigma ^{2}\left( u\right) du,
\end{gather}
is the standard Black-Scholes formula. Hence cap/floor options prices are given at $t\leq T_{0}$ by
\begin{align}
& \text{\textbf{CF}}\left( t;\mathbf{T},\mathbf{K},\mathbf{\omega },\mathbf{N}\right) =\sum\limits_{i=1}^{n}\text{\textbf{cf}}\left(T_{i};T_{i-1},T_{i},K_{i},\mathbf{\omega }_{i}\mathbf{,}N_{i}\right)  \notag
\\
& =\sum\limits_{i=1}^{n}N_{i}P_{d}\left( t,T_{i}\right) \tau_{f}\left(
T_{i-1},T_{i}\right)  \notag \\
& \times Bl\left[ F_{f}\left( t;T_{i-1},T_{i}\right) QA_{fd}\left(
t,T_{i-1},\sigma_{f,i},\sigma_{X,i}\rho_{fX,i}\right) ,K_{i},\mu
_{f,i},\sigma _{f,i},\omega _{i}\right] ,
\end{align}

Finally, for swaptions on a $T_{0}$-spot swap rate with payoff at maturity $T_{0}$ given by
\begin{equation}
\text{\textbf{Swaption}}\left( T_{0};\mathbf{T},\mathbf{S},K,N\right)
= N\text{Max}\left[ \mathbf{\omega }\left( S_{f}\left( T_{0},\mathbf{T},\mathbf{S}\right) -K\right) \right] A_{d}\left( T_{0},\mathbf{S}\right) ,
\end{equation}
the standard market-like pricing expression at time $t\leq T_{0}$, using the
discounting swap measure $Q_{d}^{\mathbf{S}}$ associated to the numeraire
$A_{d}\left( t,\mathbf{S}\right) $ on curve $\complement_{d}$,
is modified as follows
\begin{multline}
\text{\textbf{Swaption}}\left( t;\mathbf{T},\mathbf{S},K,N\right)
= N A_{d}\left( t,\mathbf{S}\right) \mathbb{E}_{t}^{Q_{d}^{\mathbf{S}}}\left\{
\text{Max}\left[ \mathbf{\omega }\left( S_{f}\left( T_{0},\mathbf{T},\mathbf{S}\right) -K\right) \right] \right\} \\
= N A_{d}\left(t,\mathbf{S}\right) Bl\left[ S_{f}\left( t,\mathbf{T},\mathbf{S}\right) QA_{fd}\left(t,\mathbf{T},\mathbf{S},\nu_{f},\nu_{Y},\rho_{fY}\right),K,\lambda_{f},\nu_{f},\omega \right].
\end{multline}
where we have used eq. \ref{QuantoAdjSwap} and the quanto adjustment term
$QA_{fd}\left( t,\mathbf{T},\mathbf{S},\nu _{f},\nu _{Y},\rho _{fY}\right)$
is given by eq. \ref{QuantoExpSwap}.

When two or more different underlying interest-rates are present, pricing expressions may become more involved. An example is the spread option, for which the reader can refer to, e.g., ch. 14.5.1 in ref. \cite{BriMer06}.

\begin{figure}[tb]
\centering
\includegraphics[scale=0.8]{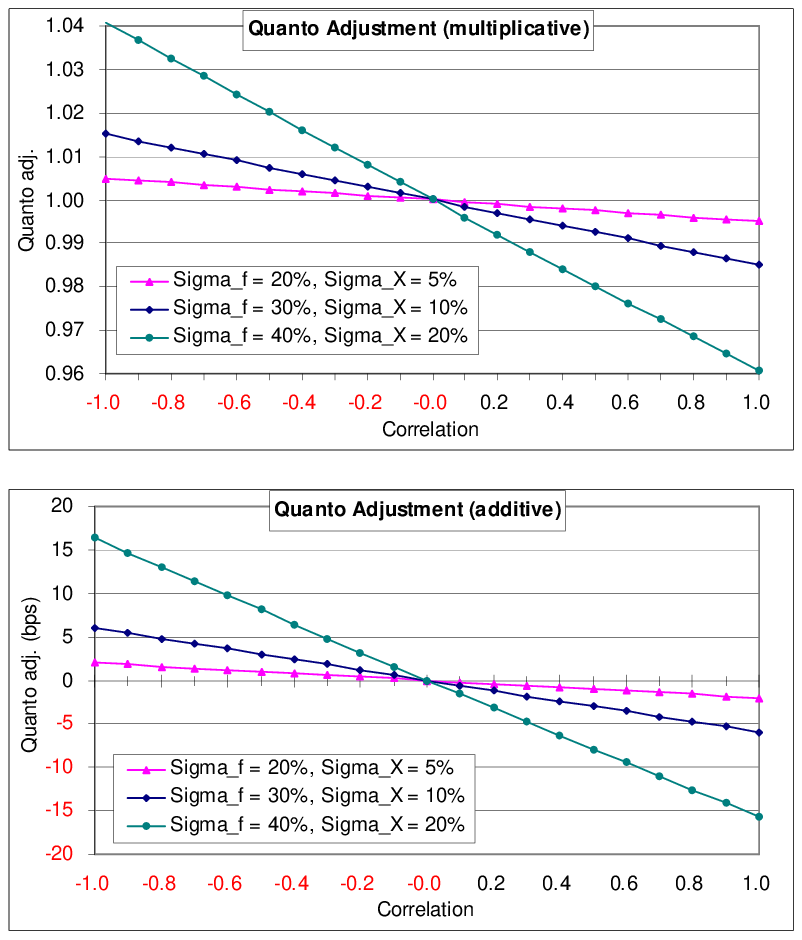}
%\vspace{-0.4cm}
\caption{Numerical scenarios for the
quanto adjustment. Upper panel: multiplicative (from eq. \protect\ref{QuantoExp}); lower panel: additive (from eq. \protect\ref{QuantoAdd}). In each figure we show the quanto adjustment corresponding to three different combinations of (flat) volatility values as a function of the correlation.
The time interval is fixed to $T_{1}-t=0.5$ and the forward rate entering eq. \protect\ref{QuantoAdd} to $4\%$, a typical value in fig. \protect\ref{FigYieldCurves}. We see that, for realistic values of volatilities and correlation, the magnitude of the adjustment may be important.}
\label{FigQuantoAdj}
\end{figure}

The calculations above show that also basic interest rate derivatives prices
include a quanto adjustment and are thus volatility and correlation
dependent.
The volatilities and the correlation in eqs. \ref{DriftAdjFwd} and \ref{DriftAdjSwap} can be inferred from market data. In the EUR market the volatilities $\sigma_f$ and $\nu_f$ can be extracted from quoted caps/floors/swaptions on Euribor6M, while for $\sigma_X$, $\rho_{fX}$ and $\nu_Y$, $\rho_{fY}$ one must resort to historical estimates.
Conversely, given a forward basis term structure, such that in fig. \ref{FigBasisAdj}, one could take $\sigma_f$ and $\nu_f$ from the market options, assume for simplicity
$\rho_{fX}\simeq \rho_{fY} \simeq 1$ (or any other functional form), and bootstrap out a term structure for the forward exchange rate volatilities $\sigma_X$ and $\nu_Y$. Notice that in this way one is also able to compare information about the internal dynamics of different market sub-areas.
\par
In fig. \ref{FigQuantoAdj} we show some numerical scenario for
the quanto adjustment in eqs. \ref{QuantoExp}, \ref{QuantoAdd}. We see that,
for realistic values of volatilities and correlation, the magnitude of the
additive adjustment may be non negligible, ranging from a few basis points
up to over 10 basis points. Time intervals longer than the 6M period used in
fig. \ref{FigQuantoAdj} further increase the effect. Notice that positive
correlation implies negative adjustment, thus lowering the forward rates
that enters the pricing formulas above.
Through historical estimation of parameters we obtain, using eq. \ref{FwdExchRate} with the same yield curves as in fig. \ref{FigBasisAdj} and considering one year of backward data, forward exchange rate volatilities below 5\%-10\% and correlations within the range $\left[-0.6;+0.4\right]$.
\par
Pricing interest rate derivatives without the quanto adjustment thus leaves,
in principle, the door open to arbitrage opportunities. In practice the
correction depends on financial variables presently not quoted on the
market, making virtually impossible to set up arbitrage positions and lock
today expected future positive gains. Obviously one may bet on his/her
personal views of future realizations of volatilities and correlation.

\subsection{\label{SecHedging}Hedging}
Hedging within the multi-curve framework implies taking into account multiple
bootstrapping and hedging instruments. We assume to have a portfolio $\Pi$ filled with a variety of interest rate derivatives with different underlying rate tenors. The first issue is how to calculate the delta sensitivity of $\Pi $. In principle, the answer is straightforward: having recognized interest-rates with different tenors as different underlyings, and having
constructed $N_C$ multiple yield curves $\mathbf{\complement }=\left\{ \complement_{d},\complement_{f}^{1},...,\complement_{f}^{N}\right\}$ using
homogeneous market instruments, we must coherently calculate the sensitivity
with respect to the corresponding market rates $\mathbf{r}^{B}=\left\{\mathbf{r}_d^B,\mathbf{r}_{f_1}^B,...,\mathbf{r}_{f_N}^B \right\}$
of \emph{each} bootstrapping instrument of \emph{each} curve\footnote{with the obvious caveat of avoiding double counting of those instruments
eventually appearing in more than one curve (3M Futures for instance could
appear both in $\complement_{d}$ and in $\complement_{f}^{3M}$curves).},
\begin{equation}
\Delta ^{B}\left( t,\mathbf{r}^{B}\right)
=\sum\limits_{i=1}^{N_{B}}\Delta_{i}^{B}\left( t,\mathbf{r}^{B}\right)
=\sum\limits_{k=1}^{N_{C}}\sum\limits_{j=1}^{N_{k}^r}
\frac{\partial \Pi \left(t,\mathbf{r}^{B}\right) }{\partial r_{j,k}^{B}},
\label{Delta}
\end{equation}
where $N_B$ is the total number of independent bootstrapping market rates, $N_k^r$ is the number of independent market rates in curve $\mathbf{\complement_k}$, $r_{j,k}^{B}$ is the \emph{j-th} independent bootstrapping market rate of curve $\mathbf{\complement_k}$ and $\Pi \left( t,\mathbf{r}^{B}\right) $ is the price at time $t$ of the portfolio $\Pi.$
\par
In practice this can be computationally cumbersome, given the high number of market instruments involved. Furthermore, second order delta sensitivities appear, due to the multiple curve bootstrapping described in sec. \ref{SecMultipleCurve}. In particular, the forwarding curves
$\left\{\complement_{f}^{1},...,\complement_{f}^{N}\right\}$ depend directly on their corresponding bootstrapping market instruments
$\left\{\mathbf{r}_{f_1}^B,...,\mathbf{r}_{f_N}^B \right\}$
but also indirectly on the discounting curve $\complement_{d}$, as
\begin{align}
\Delta ^{B}\left( t,\mathbf{r}^{B}_d\right)
& =\sum\limits_{j=1}^{N_{d}^r} \sum\limits_{\alpha=1}^{N_{d}^z}
\frac{\partial \Pi \left(t,\mathbf{r}^{B}\right) }{\partial z_{d,\alpha}^{B}}
\frac{\partial z_{d,\alpha}^{B} }{\partial r_{d,j}^{B}} \notag \\
& +\sum\limits_{j=1}^{N_{f}^r} \sum\limits_{\alpha=1}^{N_{f}^z}
\frac{\partial \Pi \left(t,\mathbf{r}^{B}\right) }{\partial z_{f,\alpha}^{B}}
\left(
\frac{\partial z_{f,\alpha}^{B} }{\partial r_{f,j}^{B}}
+ \frac{\partial z_{f,\alpha}^{B} }{\partial r_{d,j}^{B}}
\right),
\end{align}
where $\mathbf{z}^{B}=\left\{\mathbf{z}_d^B,\mathbf{z}_{f_1}^B,...,\mathbf{z}_{f_N}^B \right\}$ is the vector of $\left\{N_d^z,N_{f_1}^z,...,N_{f_N}^z \right\}$ zero rate pillars in the curves.
\par
Once the delta sensitivity of the portfolio is known for each pillar of each
relevant curve, the next issues of hedging are the choice of the set $%
\mathbf{H}$ of hedging instruments and the calculation of the corresponding
hedge ratios $\mathbf{h}$. In principle, there are two alternatives: a)\ the
set $\mathbf{H}$ of hedging instruments exactly overlaps the set $\mathbf{B}$
of bootstrapping instruments $\left( \mathbf{H\equiv B}\right) $; or, b) it
is a subset restricted to the most liquid bootstrapping instruments $\left(
\mathbf{H\subset B}\right) $. The first choice allows for a straightforward
calculation of hedge ratios and representation of the delta risk
distribution of the portfolio. But, in practice, people prefer to hedge
using the most liquid instruments, both for better confidence in their
market prices and for reducing the cost of hedging. Hence the second
strategy generally prevails. In this case the calculation of hedge ratios
requires a three-step procedure: first, the sensitivity $\mathbf{\Delta }%
^{B}=\left\{ \Delta _{1}^{B},...,\Delta _{N_{B}}^{B}\right\} $ is calculated
as in eq. \ref{Delta} on the basis $\mathbf{B}$ of all bootstrapping
instruments; second, $\mathbf{\Delta }^{B}$ is projected onto the basis $%
\mathbf{H}$ of hedging instruments\footnote{%
in practice $\mathbf{\Delta }^{H}$ is obtained by aggregating the components
$\mathbf{\Delta }_{i}^{B}$ through appropriate mapping rules.},
characterized by market rates $\mathbf{r}^{H}=\left\{
r_{1}^{H},...,r_{N_{H}}^{H}\right\} $, thus obtaining the components $%
\mathbf{\Delta }^{H}=\left\{ \Delta _{1}^{H},...,\Delta _{N_{H}}^{H}\right\}
$ with the constrain%
\begin{equation}
\Delta ^{B}=\sum\limits_{i=1}^{N_{B}}\Delta
_{i}^{B}=\sum\limits_{j=1}^{N_{H}}\Delta _{j}^{H}=\Delta ^{H};
\end{equation}%
then, hedge ratios $\mathbf{h}=\left\{ h_{1},...,h_{N_{H}}\right\} $ are
calculated as
\begin{equation}
h_{j}=\frac{\Delta _{j}^{H}}{\delta _{j}^{H}},
\end{equation}%
where $\mathbf{\delta }^{H}=\left\{ \delta _{1}^{H},...,\delta
_{N_{H}}^{H}\right\} $ is the delta sensitivity of the hedging instruments.
The disadvantage of this second choice is, clearly, that some risk - the
basis risk in particular - is only partially hedged; hence, a particular
care is required in the choice of the hedging instruments.

A final issue regards portfolio management. In principle one could keep all
the interest rate derivatives together in a single portfolio, pricing each
one with its appropriate forwarding curve, discounting all cash flows with
the same discounting curve, and hedging using the preferred choice described
above. An alternative is the segregation of homogeneous contracts (with the
same underlying interest rate index) into dedicated sub-portfolios, each
managed with its appropriate curves and hedging techniques. The
(eventually)\ remaining non-homogeneous instruments (those not separable in
pieces depending on a single underlying) can be redistributed in the
portfolios above according to their prevailing underlying (if any), or put
in other isolated portfolios, to be handled with special care. The main
advantage of this second approach is to \textquotedblleft clean
up\textquotedblright\ the trading books, \textquotedblleft
cornering\textquotedblright\ the more complex deals in a controlled way, and
to allow a clearer and self-consistent representation of the sensitivities
to the different underlyings, and in particular of the basis risk of each
sub-portfolio, thus allowing for a cleaner hedging.

\section{\label{SecCtpRisk}No Arbitrage and Counterparty Risk}
Both the forward basis and the quanto adjustment discussed in sections \ref{SecNoArbBasisAdj}, \ref{SecQuantoAdj} above find a simple financial explanation in terms of counterparty risk. From this point of view we may identify $P_{d}\left(t,T\right)$ with a default free zero coupon bond and $P_{f}\left(t,T\right)$ with a risky zero coupon bond with recovery rate $R_f$, emitted by a generic interbank counterparty subject to default risk. The associated risk free and risky Xibor rates, $L_{d}\left(T_{1},T_{2}\right)$ and $L_{f}\left(T_{1},T_{2}\right)$, respectively, are the underlyings of the corresponding derivatives, e.g. FRA$_d$ and FRA$_f$.
Adapting the simple credit model proposed in ref. \cite{Mer09} we may write, using our notation,
\begin{gather}
P_{f}\left(t,T\right) = P_{d}\left(t,T\right)R\left(t;t,T,R_f\right),
\label{riskyZCB} \\
R\left(t;T_1,T_2,R_f\right) := R_f+\left(1-R_f\right)
\mathbb{E}_t^{Q_d}\left[q_d(T_1,T_2)\right],
\label{riskyCtp}
\end{gather}
where
$q_{d}(t,T)=\mathbb{E}_{t}^{Q_{d}}\left[ 1_{\tau \left( t\right) >T}\right] $
is the counterparty default probability after time $T$ expected at time $t$ under the risk neutral discounting measure $Q_{d}$. Using eqs. \ref{riskyZCB}, \ref{riskyCtp} we may express the risky Xibor spot and forward rates as
\begin{eqnarray}
L_{f}\left( T_{1},T_{2}\right) &=&\frac{1}{\tau_{f}\left(
T_{1},T_{2}\right) }\left[ \frac{1}{P_{f}\left( T_{1},T_{2}\right) }-1\right]
\notag \\
&=&\frac{1}{\tau_{f}\left( T_{1},T_{2}\right) }\left[ \frac{1}{P_{d}\left(
T_{1},T_{2}\right) }\frac{1}{R\left(T_1;T_1,T_2,R_f\right)}-1\right] , \\
F_f\left(t;T_1,T_2\right) &=& \frac{1}{\tau_f\left(T_{1},T_{2}\right)}
\left[\frac{P_f\left(t,T_1\right)}{P_f\left(t,T_2\right)} -1 \right] \notag \\
&=& \frac{1}{\tau_f\left(T_{1},T_{2}\right)}
\left[\frac{P_d\left(t,T_1\right)}{P_d\left(t,T_2\right)}
\frac{R\left(t;t,T_1,R_f\right)}{R\left(t;t,T_2,R_f\right)} -1 \right],
\label{riskyFRA}
\end{eqnarray}
and the risky FRA$_{f}$ price at time $t$ as\footnote{in particular, contrary to ref. \cite{Mer09}, we use here the FRA definition of eq. \ref{FRAPrice}, leading to eq. \ref{FRAPriceCtp}.}
\begin{equation}
\text{\textbf{FRA}}_f\left(t;T_1,T_2,K\right) =
\frac{P_{d}\left(t,T_1\right)}{R\left(t;T_1,T_2,R_f\right)} - P_{d}\left(t,T_2\right)
\left[1+K\tau_{f}\left(T_1,T_2\right)\right], \label{FRAPriceCtp}
\end{equation}
Introducing eq. \ref{riskyFRA} in eqs. \ref{FwdBasisAdj2} and \ref{FwdBasisAdj3} we obtain the following expressions for the forward basis
\begin{eqnarray}
BA_{fd}\left( t;T_{1},T_{2}\right) &=& \frac{P_d\left(t,T_1\right)R\left(t;t,T_1,R_f\right)-P_d\left(t,T_2\right)R\left(t;t,T_2,R_f\right)}
{\left[P_d\left(t,T_1\right)-P_d\left(t,T_2\right)\right]R\left(t;t,T_2,R_f\right)}, \\
BA_{fd}^{\prime }\left( t;T_{1},T_{2}\right)
&=& \frac{1}{\tau_{d}\left(T_{1},T_{2}\right)}
\frac{P_{d}\left( t,T_{1}\right) }{P_{d}\left(t,T_{2}\right)}
\left[\frac{R\left(t;t,T_1,R_f\right)}{R\left(t;t,T_2,R_f\right)}-1\right].
\end{eqnarray}
From the FRA pricing expression eq. \ref{FRAPrice2} we may also obtain an expression for the FRA quanto adjustment
\begin{eqnarray}
QA_{fd}\left(t;T_{1},T_{2}\right)&=&
\frac{P_{d}\left(t,T_1\right)\frac{1}{R\left(t;T_1,T_2,R_f\right)} - P_{d}\left(t,T_2\right)}
{P_{d}\left(t,T_1\right)\frac{R\left(t;t,T_1,R_f\right)}{R\left(t;t,T_2,R_f\right)} - P_{d}\left(t,T_2\right)}
,\\
QA_{fd}^{\prime}\left(t;T_{1},T_{2}\right)
&=& \frac{1}{\tau_f\left(T_{1},T_{2}\right)}
\frac{P_{d}\left(t,T_1\right)}{P_{d}\left(t,T_2\right)}
\left[\frac{1}{R\left(t;T_1,T_2,R_f\right)}- \frac{R\left(t;t,T_1,R_f\right)}{R\left(t;t,T_2,R_f\right)}\right].
\end{eqnarray}
Thus the forward basis and the quanto adjustment can be expressed, under simple credit assumptions, in terms of risk free zero coupon bonds, survival probability and recovery rate. A more complex credit model, as e.g. in ref. \cite{Mor09}, would also be able to express the spot exchange rate in eq. \ref{FwdExchRate} in terms of credit variables.
Notice that the single-curve case $\mathcal{C}_{d}=\mathcal{C}_{f}$ is recovered for vanishing default risk (full recovery).

\section{\label{SecConclusions}Conclusions}
We have discussed how the liquidity crisis and the resulting changes in the
market quotations, in particular the very high basis swap spreads, have
forced the market practice to evolve the standard procedure adopted for
pricing and hedging single-currency interest rate derivatives. The new
double-curve framework involves the bootstrapping of multiple yield curves
using separated sets of vanilla interest rate instruments homogeneous in the
underlying rate (typically with 1M, 3M, 6M, 12M tenors). Prices,
sensitivities and hedge ratios of interest rate derivatives on a given
underlying rate tenor are calculated using the corresponding forward curve
with the same tenor, plus a second distinct curve for discount factors.
\par
We have shown that the old, well-known, standard single-curve no arbitrage
relations are no longer valid and can be recovered with the introduction of
a forward basis, for which simple statical expressions are given in eqs. \ref{FwdBasisAdj2}-\ref{FwdBasisAdj3} in terms of discount factors from the two
curves. Our numerical results have shown that the forward basis curves, in
particular in a realistic stressed market situation, may display an
oscillating term structure, not present in the smooth and monotonic basis
swaps market quotes and more complex than that of the discount and forward
curves. Such richer micro-term structure is caused by amplification effects
of small local differences between the discount and forwarding curves and
constitutes both a very sensitive test of the quality of the bootstrapping
procedure (interpolation in particular), and an indicator of the tiny, but
observable, differences between different interest rate market areas. Both
of these causes may have appreciable effects on the price of interest rate
instruments, in particular when one switches from the single-curve towards
the double-curve framework.
\par
Recurring to the foreign-currency analogy we have also been able to
recompute the no arbitrage double-curve-single-currency market-like pricing
formulas for basic interest rate derivatives, zero coupon bonds, FRA, swaps
caps/floors and swaptions in particular. Such prices depend on forward or
swap rates on curve $\complement_{f}$ corrected with the well-known quanto
adjustment typical of cross-currency derivatives, naturally arising from the
change between the\ numeraires, or probability measures, naturally
associated to the two yield curves. The quanto adjustment depends on the
volatility $\sigma _{f}$ of the forward rates $F_{f}$ on $\complement_{f}$,
of the volatility $\sigma _{X}$ of the forward\ exchange rate $X_{fd}$
between $\complement_{f}$ and $\complement_{d}$, and of the correlation $%
\rho _{fX}$ between $F_{f}$ and $X_{fd}$. In particular, a non-trivial
adjustment is obtained if and only if the forward exchange rates $X_{fd}$
are stochastic ($\sigma _{X}\neq 0$) and correlated to the forward rate $%
F_{f}$ ($\rho _{fX}\neq 0$). Analogous considerations hold for the swap rate
quanto adjustment.
Numerical scenarios show that the quanto adjustment can be important, depending on volatilities and correlation. Unadjusted interest rate derivatives' prices are thus, in principle, not arbitrage free, but, in practice, at the moment the market does not trade enough instruments to set up arbitrage positions.
\par
Finally, both the forward basis and the quanto adjustment find a natural financial explanation in terms of counterparty risk within a simple credit model including a default free and a risky zero coupon bond.
\par
Besides the lack of information about volatility and correlation, the present framework has the advantage of introducing a minimal set of parameters with a transparent financial interpretation and leading to familiar pricing formulas, thus constituting a simple and easy-to-use tool for practitioners and traders to promptly intercept possible market evolutions.

\bibliographystyle{alpha}
\bibliography{FinanceBibliography}

\end{document}